# Choice of adaptive sampling strategy impacts state discovery, transition probabilities, and the apparent mechanism of conformational changes

Maxwell I. Zimmerman[1], Justin R. Porter[1], Xianqiang Sun[1], Roseane R. Silva[1], and Gregory R. Bowman[1,2,3,*]

[1]Department of Biochemistry and Molecular Biophysics, Washington University School of Medicine, St. Louis, Missouri 63110, United States

[2]Department of Biomedical Engineering, and [3]Center for Biological Systems Engineering, Washington University in St. Louis, St. Louis, Missouri, 63110, United States

*Corresponding Author: bowman@biochem.wustl.edu

**Abstract**

Interest in equilibrium-based sampling methods has grown with recent advances in computational hardware and Markov state modeling (MSM) methods, yet outstanding questions remain that hinder widespread adoption. Namely, how do sampling strategies explore conformational space and how might this influence predictions? Here, we seek to answer these questions for four commonly used sampling methods: 1) a long simulation, 2) many short simulations, 3) adaptive sampling, and 4) FAST. We first develop a theoretical framework for analytically calculating the probability of discovering states and uncover the drastic effects of varying the number and length of simulations. We then use kinetic Monte Carlo simulations on a variety of physically inspired landscapes to characterize state discovery and transition pathways. Consistently, we find that FAST simulations discover target states with the highest probability and traverse realistic pathways. Furthermore, we uncover the pathology that short parallel simulations sometimes predict an incorrect transition pathway by crossing large energy barriers that long simulations would typically circumnavigate, which we refer to as "pathway tunneling". To protect against tunneling, we introduce "FAST-string", which samples along the highest-flux transition paths to refine an MSMs transition probabilities and discriminate between competing pathways. Additionally, we compare MSM estimators in describing thermodynamics and kinetics. For adaptive sampling, we recommend normalizing the transition counts out of each state after adding pseudo-counts to avoid creating sources or sinks. Lastly, we evaluate our insights from simple landscapes with all-atom molecular dynamics simulations of the folding of the λ-repressor protein. We find that FAST-contacts predicts the same folding pathway as long simulations but with orders of magnitude less simulation time.

**Introduction**

The use of all-atom molecular dynamics (MD) simulations for long time-scale phenomena are often thwarted by insufficient computational resources. Many interesting biological processes occur on the millisecond to second timescale, where a single simulation may take longer than a lifetime to gather. Notable attempts to alleviate hardware limitations are the purpose-built ANTON supercomputers. These supercomputers are an engineering feat, yet are still sampling limited and not accessible to many researchers. Since increasing commodity hardware performance by many orders of magnitude is not likely in the immediate future, the observation of interesting biological phenomena requires the use of clever sampling techniques.

A common technique to increase the observation of long time-scale phenomena is to alter the underlying energy landscape. These methods aim to guide a simulation towards some end goal or toward the exploration of a set of order parameters. Some examples include, Gō models,[1,2] replica-exchange,[3-5] steered MD,[6,7] accelerated MD,[8,9] meta-dynamics,[10,11] among others.[12-14] Unfortunately, these methods do not capture proper kinetic information, and can traverse unrealistically high energy barriers. Here, we are particularly interested in sampling methods that access long time-scale phenomena without perturbing the underlying energy landscape, such that both thermodynamic and kinetic properties can be inferred.

As an alternative to a single long simulation, many independent simulations can be run in parallel. Combined, these parallel simulations tractably capture time-scales longer than any single simulation. To illustrate: if we assume that the transition between conformational states A and B follows a Poisson process, the probability of observing a transition to state B is dependent only on the aggregate simulation time from A, not the length of each simulation.[15] Put another way, the probability of traversing a single energy barrier is based on the number of attempts to cross that barrier, regardless of whether they are in parallel or successive. Thus, parallel simulations may offer a significant enhancement in the observation of rare events, since it is usually easier to add more computational resources than to make them faster. This is the strategy of Folding@home, which takes advantage of around 100,000 personal computers, whose resources are donated for massively distributed MD simulations.[16]

For large sets of independent simulations that are in local equilibrium (i.e. they sample from the underlying energy distribution), we can reconstruct both the proper thermodynamics and kinetics with the use of Markov state models (MSMs).[17] An MSM is a network model that describes a protein's energy landscape in terms of a set of structural states the protein tends to adopt and the probabilities of transitioning between neighboring states in a fixed time interval. The utility of an MSM depends on accurately estimating the conditional transition probabilities between conformational states, without requiring that any individual simulation achieve global equilibration. As a consequence, the number of times different states are sampled does not need to be Boltzmann distributed for an accurate description of their populations at equilibrium, provided that estimates of transition probabilities are accurate. MSMs have recently succeeded in guiding the design of new proteins[18,19] and allosteric modulators,[20] among many other applications.[17,21-25]

MSMs' ability to integrate information from many parallel simulations whose starting states are not necessarily Boltzmann distributed opens the possibility of performing adaptive sampling. First developed for refining MSMs by identifying conformational states that contribute the most to statistical uncertainty,[26] adaptive sampling schemes typically have the following steps: 1) run simulations, 2) build an MSM from simulations, 3) rank each state by some metric, 4) start new simulations from the highest ranked states, and 5) repeat steps 2-4 for some number of rounds or until a convergence criterion is met. The main difference between adaptive sampling algorithms is in the metric for ranking and selecting states for future sampling.[26-32] Recently, we have developed the goal-oriented sampling algorithm, Fluctuation Amplification of Specific Traits (FAST), that ranks states on some structural metric in addition to a statistical metric.[33,34] We have demonstrated that this method increases the rate of state exploration by at least an order of magnitude, and additionally, can capture thermodynamic and kinetic information that agrees with a multitude of experiments.[19,35]

Each of the equilibrium-based sampling methods mentioned above (long-, parallel-, adaptive-, and FAST-simulations) should converge on identical MSMs, provided with near infinite sampling. Unfortunately, for most systems of interest, simulations are not able to reach global equilibrium, and are usually significantly under-sampled. It should be noted that FAST, and other adaptive sampling algorithms, do not *increase* the amount of sampling, but rather focus sampling efforts to specific regions of conformational space to make the most of limited computational resources. With that, the functional differences between methods are simply the rates at which specific sections of conformational-space are explored. However, it is not completely understood how each of these methods influences the probability of discovering states, nor how this influences the mechanism of conformational changes that is observed, especially when conformational sampling is far from global equilibrium.

In this work, we seek to assess the relative performance of different sampling strategies. We develop an analytical expression for the probability of discovering a conformational state for very simple landscapes. We find that state discovery is dependent on the number and length of simulations, in addition to the shape of the energy landscape. We then examine the performance of the four equilibrium-based sampling methods above in finding the highest-flux pathway between two states, for a variety of energy landscapes. These results are very informative for tuning the many hyperparameters in adaptive sampling, and even identify pitfalls that should be avoided. Lastly, we demonstrate that insights from our simple landscapes are consistent with observations using all-atom MD simulations, by generating folding trajectories of a fast-folding version of the λ-repressor.

**Theory**

To understand how the probability of discovering a state on a particular landscape is dependent on sampling, we develop a mathematical formalism for describing the probability that a set of simulations will discover a particular conformational state. First, we consider sampling to occur on a discretized energy landscape with $N$ conformational states, where the state index is represented as $n_i$, $i$ = 1, ..., $N$. Transitions between states are described by the $N \times N$-transition probability matrix, $T_{ij}$, which is the probability of transitioning from state $n_i$ to $n_j$ at a specified lag-time, $\tau$. A simulation on this landscape of $K$-steps is denoted with the symbol $\mathbf{X}$, where the conformation at the $k$-th time step is $X_k$, $k$ = 1, ..., $K$. For a dataset with M simulations, we denote the *m*-th simulation as $\mathbf{X}_m$, $m$ = 1, ..., $M$. For multiple simulations of various lengths

(different number of time steps), we choose **K** to represent a vector of lengths, where $K_m$, $m =$ 1, …, $M$, is the length of the $m$-th simulation.

Towards our goal of describing the probability of discovering a particular conformational state on an energy landscape given sampling parameters, we introduce the $N \times N$-matrix, $D_{ij}^{K,M}$, which indicates if state $n_j$ is ever discovered within the trajectories $\mathbf{X}_M$, started from state $n_i$ with lengths described by **K**. For example, if state $n_j$ is a state within the trajectories, $D_{ij}^{K,M}$ is 1, otherwise it is 0. This can be represented with,

$$D_{ij}^{K,M} = \begin{cases} 1 & if\ n_j \in \mathbf{X}_M \\ 0 & if\ n_j \notin \mathbf{X}_M \end{cases} \quad [1]$$

While this can be determined for a set of trajectories, we wish to know the probability of having observed a state, *a priori*, or $P(D_{ij}^{K,M} = 1)$. This is the probability of discovering state $n_j$ given the sampling parameters **K**. For short hand, we call these probabilities the "discover probabilities".

Before providing an expression for $P(D_{ij}^{K,M} = 1)$, we must first introduce another $N \times N$-matrix, $v_{ij}^k$, which indicates if the conformation at the $k$-th step of a single trajectory, **X,** belongs to the state $n_j$, when started from state $n_i$.

$$v_{ij}^k = \begin{cases} 1 & if\ X_k = n_j \\ 0 & if\ X_k \neq n_j \end{cases} \quad [2]$$

Additionally, we are interested in the probability of this event occurring, denoted as $P(v_{ij}^k = 1)$. Since only one conformation at the $k$-th step can be observed, each row of $P(v_{ij}^k = 1)$ is a normalized probability vector indicating the state index at time $k$. For the trivial case of the $0^{th}$-step (i.e. before a simulation is generated), the probability of being in the starting state is 1, and everywhere else, 0:

$$P(v_{ij}^{k=0} = 1) = I$$

Since $P(v_{ij}^k = 1)$ is a list of probability vectors, we can propagate the probabilities a time step (the lag-time, $\tau$) using the transition probability matrix.

$$P(v_{ij}^k = 1) = \begin{cases} I & if\ k = 0 \\ P(v_{ij}^{k-1} = 1)\,T & if\ k > 0 \end{cases} \quad [3]$$

This expression is useful for determining $P(D_{ij}^{K,M} = 1)$, since the probability of ever visiting a state is the complement of not visiting it at each time step. For example, the probability of discovering state $n_j$ after one step is the complement of not discovering it before and after one step:

$$P\left(D_{ij}^{K=\{1\},M=1}=1\right)=1-\left(1-P(v_{ij}^0=1)_{ij}\right)*\left(1-P(v_{ij}^1=1)_{ij}\right)=1-(1-I_{ij})*(1-$$

$$T_{ij})=\begin{cases}1 & if\ i=j\\ T_{ij} & if\ i\ne j\end{cases}$$

This reasoning holds true for a single step in a simulation, although does not for more than one step. What is required is an expression for the probability of being in a state at time step, $k$, conditional on not having discovered state $n_j$ for all of the previous steps. We represent this expression as, $P\left(v_{i'j'}^k=1\mid\{v_{ij}^{k'}=0\ \forall\ k'<k\}\right)$, which can be evaluated with the following:

$$P\left(v_{i'j'}^k=1\mid\{v_{ij}^{k'}=0\ \forall\ k'<k\}\right)=\begin{cases}I & if\ k=0\\ P\left(v_{i'j'}^{k-1}=1\mid\{v_{ij}^{k'}=0\ \forall\ k'\le(k-1)\}\right)T & if\ k>0\end{cases}$$

[4]

For each step in the recursive calculation, the $j^{th}$ column of $P\left(v_{i'j'}^{k-1}=1\right)$ is set to 0, and each row is then normalized to unity. This is described in more detail in the supporting information.

Using this definition, we can extend our expression of the discover probabilities to include an arbitrary number of steps, $K$. In a single simulation, we can see that the probability of discovering a state is:

$$P\left(D_{ij}^{K=\{K\},M=1}=1\right)=1-\prod_{k=0}^{K}\left(1-P\left(v_{i'j'}^k=1\mid\{v_{ij}^{k'}=0\ \forall\ k'<k\}\right)_{ij}\right)\qquad[5]$$

Since the probability of discovering a state within a simulation is independent of the probability in another simulation, the discover probabilities for multiple simulations is the complement of not discovering a state in any of the individual simulations. For example, in the case of two simulations with lengths $K_0$ and $K_1$,

$$P\left(D_{ij}^{K=\{K_0,K_1\},M=2}\right)=1-\left(1-P\left(D_{ij}^{K=\{K_0\},M=1}=1\right)_{ij}\right)*\left(1-P\left(D_{ij}^{K=\{K_1\},M=1}=1\right)_{ij}\right)$$

This can be generalized to an arbitrary number of simulations with arbitrary lengths:

$$P(D_{ij}^{\mathbf{K},\mathbf{M}}=1)=1-\prod_{m=1}^{M}\left[\prod_{k=0}^{K_m}\left(1-P\left(v_{i'j'}^k=1\mid\{v_{ij}^{k'}=0\ \forall\ k'<k\}\right)_{ij}\right)\right]\qquad[6]$$

This gives us our final expression for state discovery as a function of our equilibrium-sampling parameters.

**Results and Discussion**
*There are different advantages to running many short or a few long simulations*

From equation 6, it is clear that the probability of discovering a state is influenced by four parameters: 1) the number of trajectories, 2) the lengths of the trajectories, 3) the starting state, and 4) the shape of the landscape being sampled. Strikingly, this implies that the probability of discovering a state can be drastically distinct between a single long simulation and many short simulations, though this is only true for finite sampling since $P(D_{ij}^{K,M} = 1) \to 1$ as either, $M \to \infty$, or $K_m \to \infty$. It may seem tempting to seek the global optimum sampling parameters, however, sampling is strongly dependent on the specifics of the landscape itself. Additionally, different goals may necessitate different sampling strategies, i.e. is the goal to discover as many states as possible, or to discover a pathway between a particular set of states? From this, our goal is to characterize different sampling strategies for a variety of different landscapes to gain insight into their appropriate uses.

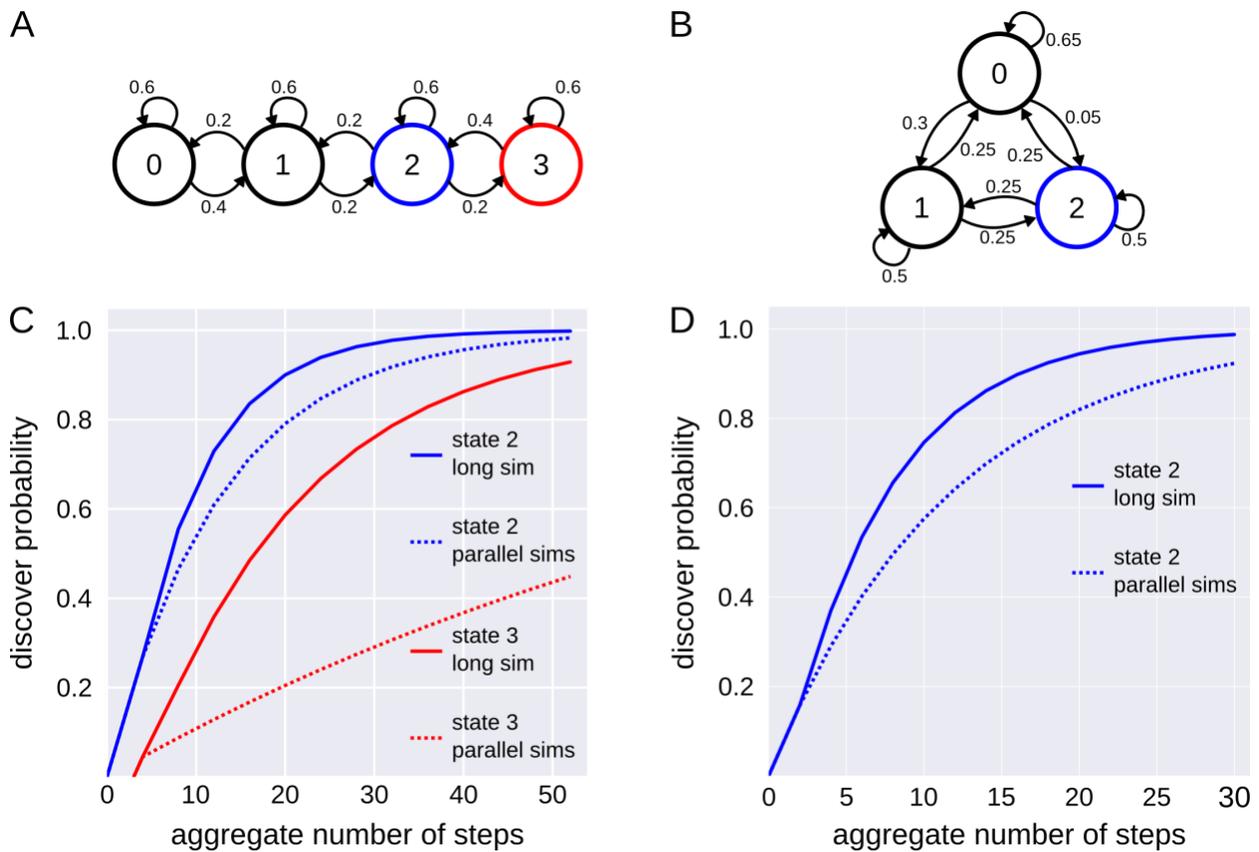

Figure 1: The probability of discovering particular states on simplified landscapes as a function of the number and length of simulations from equation 6. (A) Four states, arranged linearly, have transitions to themselves and their direct neighbors to varying degrees. States 2 and 3 are colored blue and red for visual aid. (B) A fully connected 3 state system. The probability of transitioning from state 0 to 2 is very low. (C) The probability of discovering state 2 (blue) or state 3 (red) with either a single long simulation (solid line) or many simulations consisting of 4 steps (dashed line) for the landscape in panel A. (D) The probability of discovering state 2 (blue) with either a single long simulation (solid line) or many simulations consisting of 2 steps (dashed line) for the landscape in panel B.

As a first test, we constructed a simple landscape where four states are connected in a linear arrangement (Figure 1A). Here, each state can transition to either a neighbor or itself, with differing probabilities. We imagine that these states represent a conformational landscape where each successive state is progress along some order parameter. Starting from state 0, the first state in the chain, we calculate the probability of discovering the other states from either a single trajectory or many parallel trajectories with an equivalent aggregate amount of simulation. Figure 1C depicts the probability of discovering states 2 (blue curves) and 3 (red curves) from a single trajectory at various time-steps (solid lines), or some number of parallel trajectories with 4 time-steps each (dashed lines). We see that the long simulations have a higher probability of reaching states 3 and 4 than do parallel simulations. For this shaped landscape, the discrepancy between long and parallel simulations widens with the number of states. This makes intuitive sense from equation 6, because we see that the probability of a simulation making 2 successive transitions is different than one of two simulations making 2 successive transitions.

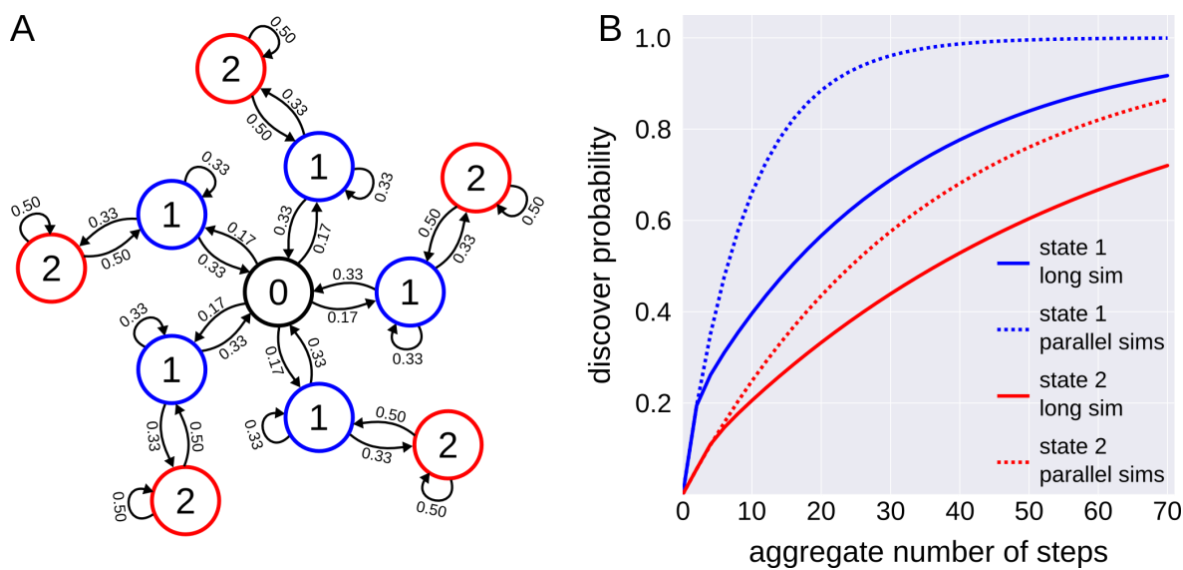

Figure 2: The probability of discovering particular states on a star-shaped landscape as a function of simulation length and number of simulations from equation 6. (A) Network representation of the star-shaped landscape. Due to symmetry in the transition probabilities, a simulation started from state 0 has equal probability of reaching any of the states labeled 1, as well as any of the states labeled 2. State 0 also has a self-transition probability of 0.17, but this edge is omitted for visual clarity. (B) The probability of discovering a particular state 1 (blue) or state 2 (red) with either a single long simulation (solid line) or parallel simulations consisting of 2 steps (dashed line).

We should note that the fully connected landscape in Figure 1B also displays this property, indicating that it is not an artifact of the way we have drawn the landscape. Here, the probability of transitioning between state 0 to 2 is very low, making the more probable route go through the transition state, 1. This leaves parallel simulations at a disadvantage of having to take the longer route to observe the transition, making this observation less probable. Interestingly, this also indicates that it is possible to consistently stumble upon an incorrect conclusion for the transition pathway; a trivial example being that many 1-step simulations started from state 0 would incorrectly predict the pathway as going directly from state 0 to 2. It is an important point that this result arises from the probability of discovering certain states, and their transitions, but not from the estimates of each states conditional transition probabilities, which should remain preserved across sampling methods. Therefore, in addition to understanding how sampling affects state discovery, we are interested in how the state discovery influences the predicted mechanism of conformational changes (e.g. the highest probability transition pathways between two sets of states). We investigate this idea in more detail in later sections.

So far, linear and fully connected landscapes might lead one to believe that long simulations are always advantageous in state discovery, but this is not the case when landscapes have entropic barriers. For many realistic systems, it is likely that a particular conformational state has many other states that it can transition to. To capture this transitional entropy, we generated the star-shaped landscape depicted in Figure 2A. This landscape has a central state and 5 arms, which is reminiscent of a "kinetic hub" where unfolded/high-energy states typically pass through the folded state to reach other unfolded/high-energy states.[36] Parallel simulations have a significantly higher probability to discover any of the states on this landscape, compared with equal aggregate time of the long simulation. We reason that the long simulations are penalized by having to backtrack to explore each of the arms, whereas the parallel simulations have a high probability of sampling multiple arms simultaneously. This effect becomes more drastic as the dimensionality of the state-space increases. Furthermore, this landscape provides a nice example that the optimal sampling scheme is strongly dependent on the shape of the landscape.

These simple landscapes provide valuable insight into how long or parallel simulations affect state discovery, setting a baseline for characterizing more complicated sampling schemes, such as adaptive sampling. Towards this goal, we generated a series of larger landscapes, which emulate common challenges in the sampling of proteins. To aid in human intuition, these landscapes are two-dimensional energy surfaces projected onto a grid, where each point on the grid represents a conformational state with a single potential energy. Each state can have up to four connected neighbors, with transitions governed by the Metropolis criterion. In the next few sections, we make use of kinetic Monte Carlo simulations on these landscapes using four different sampling methods: 1) a single long simulation (referred to as "long"), 2) many short simulations (referred to as "parallel"), 3) counts-based adaptive sampling (referred to as "counts"), and 4) our goal-oriented FAST algorithm (referred to as "FAST"). Although there are many adaptive sampling algorithms, we chose to use counts because it has been shown to be the best at indiscriminately discovering new states.[28,33] The specifics of sampling are described in greater detail in the methods section. Furthermore, we aim to characterize each method based on three different criteria: 1) ability to discover a target state, 2) ability to predict realistic transition pathways, and 3) ability to estimate accurate transition probabilities.

*FAST is most likely to discover the target state*
The first landscape that we generated was inspired by the challenge of using MD simulations to find the native state of a cooperatively folding protein. Two common tasks include: 1) to determine the native conformational state given an amino-acid sequence, also known as a structure prediction problem,[37-39] and 2) explore the preferred pathway(s) from an unfolded state to the native state.[40,41] We chose to start with one of the simplest possible models, a minimally frustrated folding-funnel (Figure 3).[42,43] Here, there is a reasonably smooth energetic gradient from a high-energy starting-state to the low-energy target state. The solid colored lines represent the three highest-flux pathways from the start to the target.

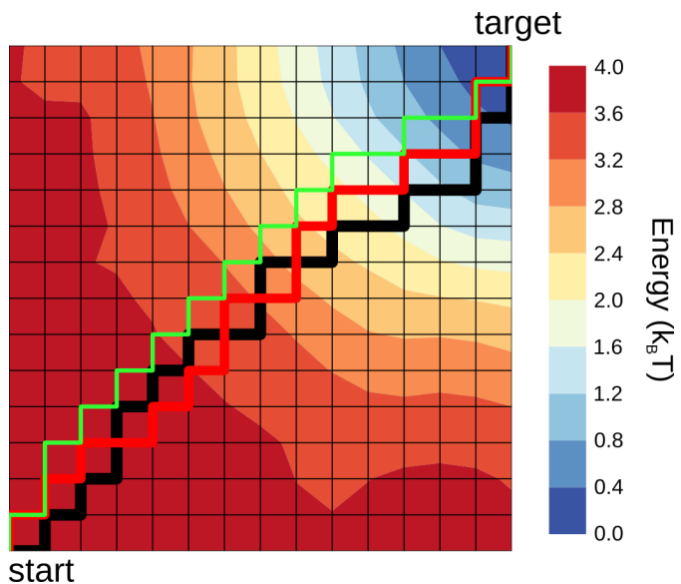

Figure 3: An energy landscape inspired by a simple folding funnel. Conformational states are located at the vertices of the grid, where the color at this point represents the energy of that state. States can have up to 4 neighbors to transition with. Solid lines (black, red, and green) indicate the three highest flux pathways from the start to the target state, where line thickness is proportional to the flux along the particular path.

To characterize state-discovery on this landscape, we performed 5,000 independent trials of each sampling method, with equivalent aggregate simulation times, as is described in the methods. We then calculate the probability of discovering a given state (which we refer to as the discover probabilities) for the four methods, by averaging the results of equation 1 for each trial. We note that we terminate the algorithm after reaching the target state, since we are mainly concerned with the initial pathway to the target; including excessive sampling after reaching the target convolutes the results with what happens afterwards. Additionally, trimming the data after discovering the end state does not affect the discover probabilities of the end state itself.

If the goal is to simply reach the end-state, FAST does so with the highest probability. The discover probabilities of the target state are 1.0 ± 7E-4, 0.94, 0.62 ± 7E-3, and 2.2e-5 for FAST, long, counts, and parallel simulations respectively (this value for long and parallel simulations come from equation 6). It is not a surprise that FAST is the best at reaching the end state, since it is the only method that uses knowledge of the end state in its sampling and we have previously shown FAST's ability to reach a target state with orders of magnitude less simulation.[33] Of greater interest here is the difference between the observation of states along the way to the target.

Towards this goal, we plot the discover probabilities for each method in Figure 4, which reveals distinct patterns for each sampling method. We find it extremely beneficial to view the discover probabilities for each state in this manner, since it provides intuition for the ways that each method explores the landscape before reaching the target. Analysis of the long simulations indicates that they have a propensity to sample around the native-well before reaching the target state. The 25 states closest to the target have over a 0.9 probability of being discovered first. Conversely, parallel simulations rarely venture near the target, but thoroughly explore the landscape around the starting state. Strikingly, this suggests that parallel simulations would require orders of magnitude more aggregate simulation time than the long simulation to reliably observe a transition to the target. In fact, this is what we observe for MD simulations of the λ-repressor in a later section.

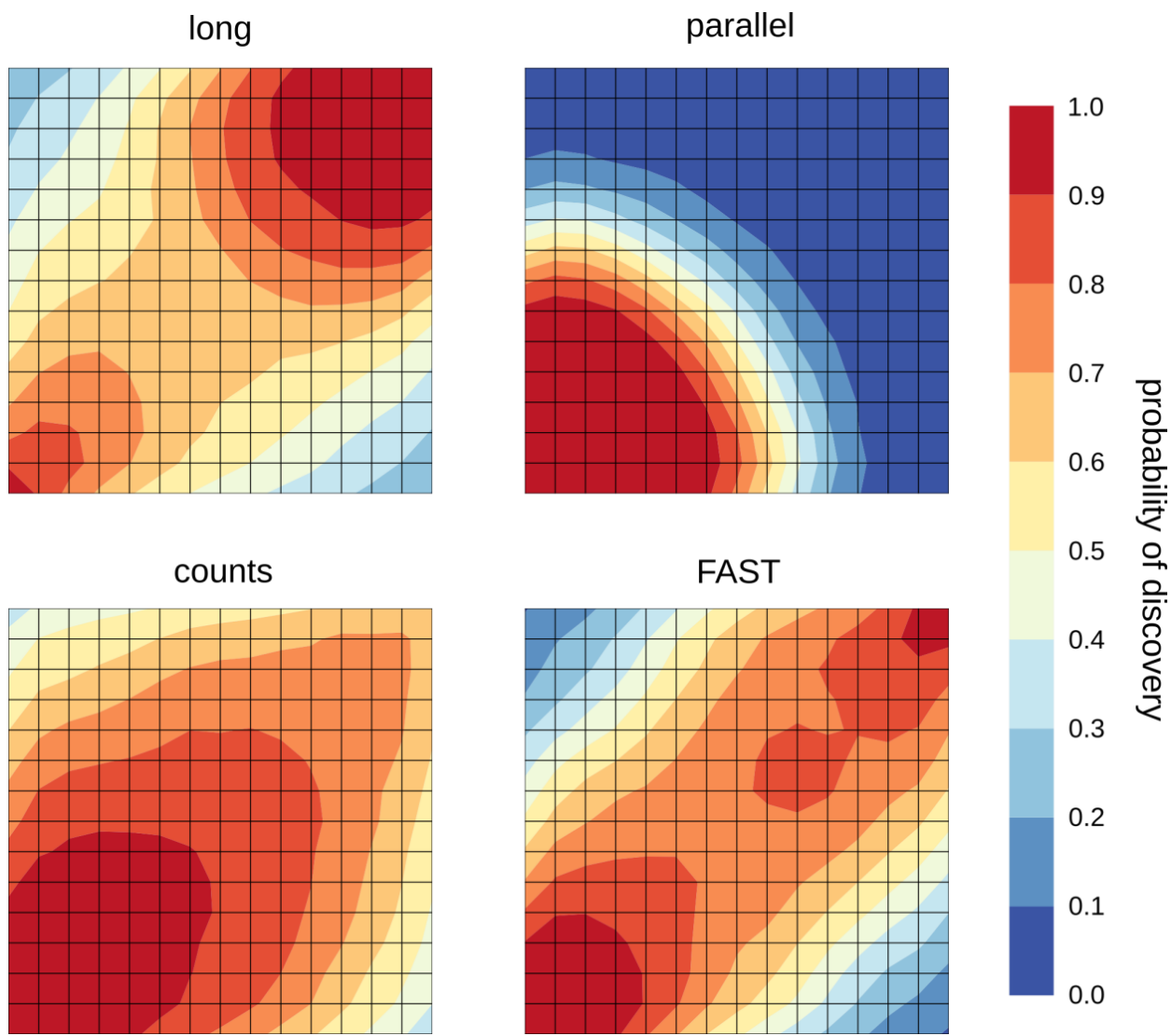

Figure 4: The discover probabilities (the probability that a simulation set observes a particular state) on the funneled landscape in Figure 3. Shown are the probabilities for four sampling strategies, a single long simulation, many parallel simulations, counts-based adaptive sampling, and the goal-oriented FAST simulations.

Unlike the other sampling strategies, counts-based adaptive sampling has an elevated propensity to explore the high-energy edges of the funneled landscape. Compared to the long simulations, counts has almost twice the probability of discovering the states furthest from the start and the target, yet, nearly half the probability of discovering the target itself. This is because counts indiscriminately discovers new states, particularly in high-energy neighborhoods where low count states are prevalent. This property enables counts-based sampling to lead in state discovery, with an average of 183.3 ± 12.3 states discovered, in comparison to 168.5 ± 12.3, 144.2 ± 24.0, and 72.7 ± 10.1 for FAST, long, and parallel simulations respectively. Interestingly, counts-based sampling's propensity to climb energy barriers actually hinders its ability to follow a simple gradient to the global minimum.

Therefore, counts-based simulations may actually be a poor choice for many applications, despite its ability to discover many states, because it will dedicate significant computational resources to sampling irrelevant (e.g. high-energy) states. On the other hand, FAST simulations are very directed.

On this funneled landscape, FAST not only has a higher probability of discovering the states along the highest-flux pathways to the global minimum, but also provides the best estimates of their transition probabilities. Using a relative entropy metric to quantify the deviation of MSMs built with each method from the true landscape, as we have done previously,[33,36] we find that FAST and long simulations have the lowest deviations for states in the top three highest-flux pathways. These relative entropies, ascending, are 0.58 ± 0.46, 0.84 ± 0.80, 1.96 ± 0.76, and 2.46 ± 5E-2 for FAST, long, counts, and parallel simulations, respectively. This result suggests that FAST matches long simulations' ability to reach distant conformations, parallel simulations' ability to thoroughly explore particular regions of conformational space, and adaptive sampling's flexibility.

Taken together, the funneled landscape provides a coarse view of each sampling methods behavior. With the perspective that aggregate simulation time is a finite resource, we can imagine the differences between sampling methods being the amount of this resource spent on each region of the conformational landscape. For example, it appears that parallel simulations spend the majority of this resource around the starting state, counts spreads it across the landscape, long simulations divvy it up proportional to neighboring states' energy, and FAST spends it on the states that optimize its objective. In this case, there are minimal barriers to prevent counts from spreading, and the states that optimize FAST's objective are nearly a straight line from the start to the target. In the following section, we add a layer of complexity to see if adaptive sampling can truly adapt to roadblocks in energy landscapes.

*Adaptive sampling navigates obstacles*

To mimic the complexities of more realistic landscapes, we generated the rugged landscape in Figure 5A. This rugged landscape provides an interesting challenge to not just discover the target state, but also discover the preferred pathways. The three highest-flux pathways between the start and the target state are shown in Figure 5A, which each require navigation around large energy barriers. As an added difficulty, there exist alternative routes around the barriers, with differing fluxes. Although sampling is stochastic, and any individual run has the potential to proceed along an arbitrary path, we expect the distribution of paths to resemble the actual highest-flux paths. Of special interest is how FAST navigates the landscape, since it strongly uses structural information in reseeding simulations. We wish to confirm that it does not cut across high-energy barriers in an effort to maximize its objective function.

Similar to the performance on the previous landscape, FAST outperforms the alternative approaches in discovering the target state. This is best seen from each methods' discover probabilities on this landscape (Figure S3), where FAST clearly has the highest probability of discovering the target state. In addition, FAST is most likely to discover the states along the actual highest-flux pathways, which suggests that FAST also predicts the correct pathways. To better quantify this, we characterize the probability that a state is predicted to be on pathway from the start to the target. This is done by calculating the highest-flux pathway for each of our 5,000 trials and determining the number of times a state is observed. Averaging this leaves us with a state value of 1 if it is always observed when transitioning from the start to the target, and 0 if it is never observed.

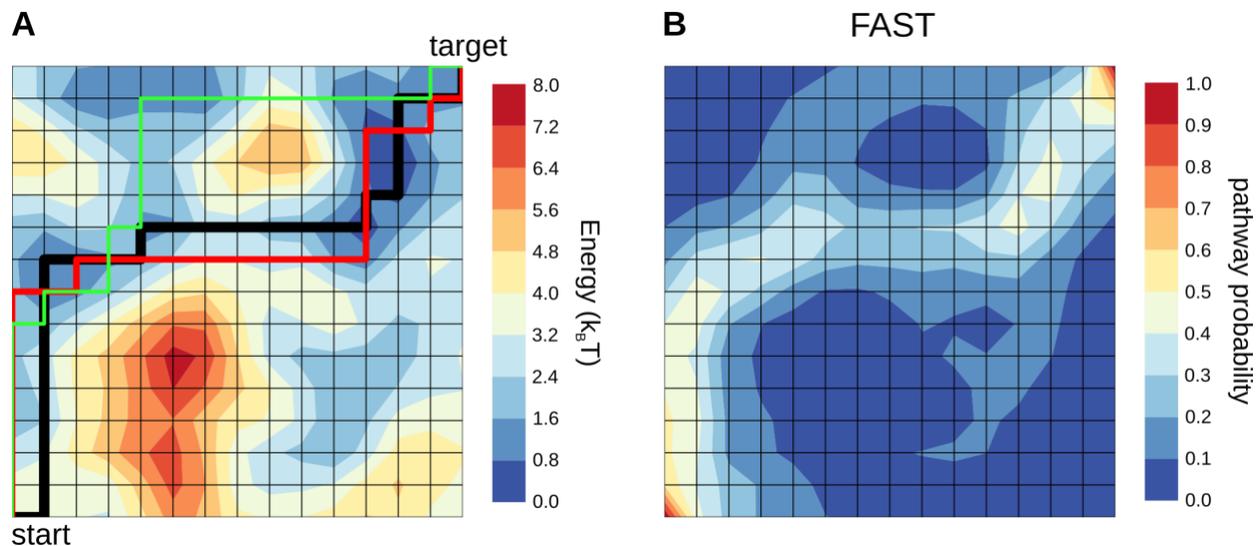

Figure 5: The performance of FAST on a rugged landscape. (A) An energy landscape inspired by a folding funnel with random obstacles. Conformational states are located at the intersection of the grid lines, where the color at this point represents the energy of that state. Solid lines (black, red, and green) indicate the three highest flux pathways from the start to the target state, where line thickness is proportional to the flux along the particular path. (B) The probability that a FAST simulation set will predict a state to be in the highest-flux path from the start to the target state.

Inspection of the pathway probabilities for FAST (Figure 5B) reveals its ability to navigate around obstacles. The predicted pathways from the start to the target do not pathologically cut across the energy barriers, but mimic the route taken by the three highest-flux pathways that were calculated from the underlying transition probabilities. Furthermore, the predicted pathways of FAST and counts resemble the predicted pathways obtained from the long simulations (Figure S4). This is consistent with the hypothesized benefits of adaptive and goal-oriented sampling: since each simulation is in local equilibrium, the probability of traversing any individual barrier remains unchanged, and thus, transitions will typically occur along realistic pathways.

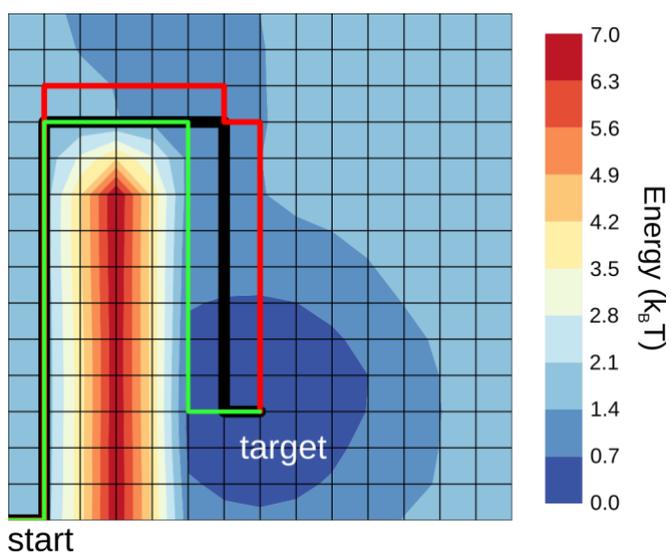

Figure 6: An energy landscape where the preferred pathway is not the shortest distance between the start and the target state. Conformational states are located at the intersection of the grid lines, where the color at this point represents the energy of that state. Solid lines (black, red, and green) indicate the three highest flux pathways from the start to the target state, where line thickness is proportional to the flux along the particular path.

*Pathway tunneling: observing an unfavorable pathway due to sampling artifacts*
The landscapes considered so far have been well suited for use with FAST, largely because the simple geometric function used in our FAST ranking (i.e. distance to the target state) is a reasonable surrogate for kinetic proximity to the target. However, there are many instances where finding a reasonable surrogate may be difficult. For example, there are many systems where transitioning between geometrically similar conformations may require partial unfolding of a protein.[44] In these cases, the optimal transition path would have, at times, unfavorable state rankings.

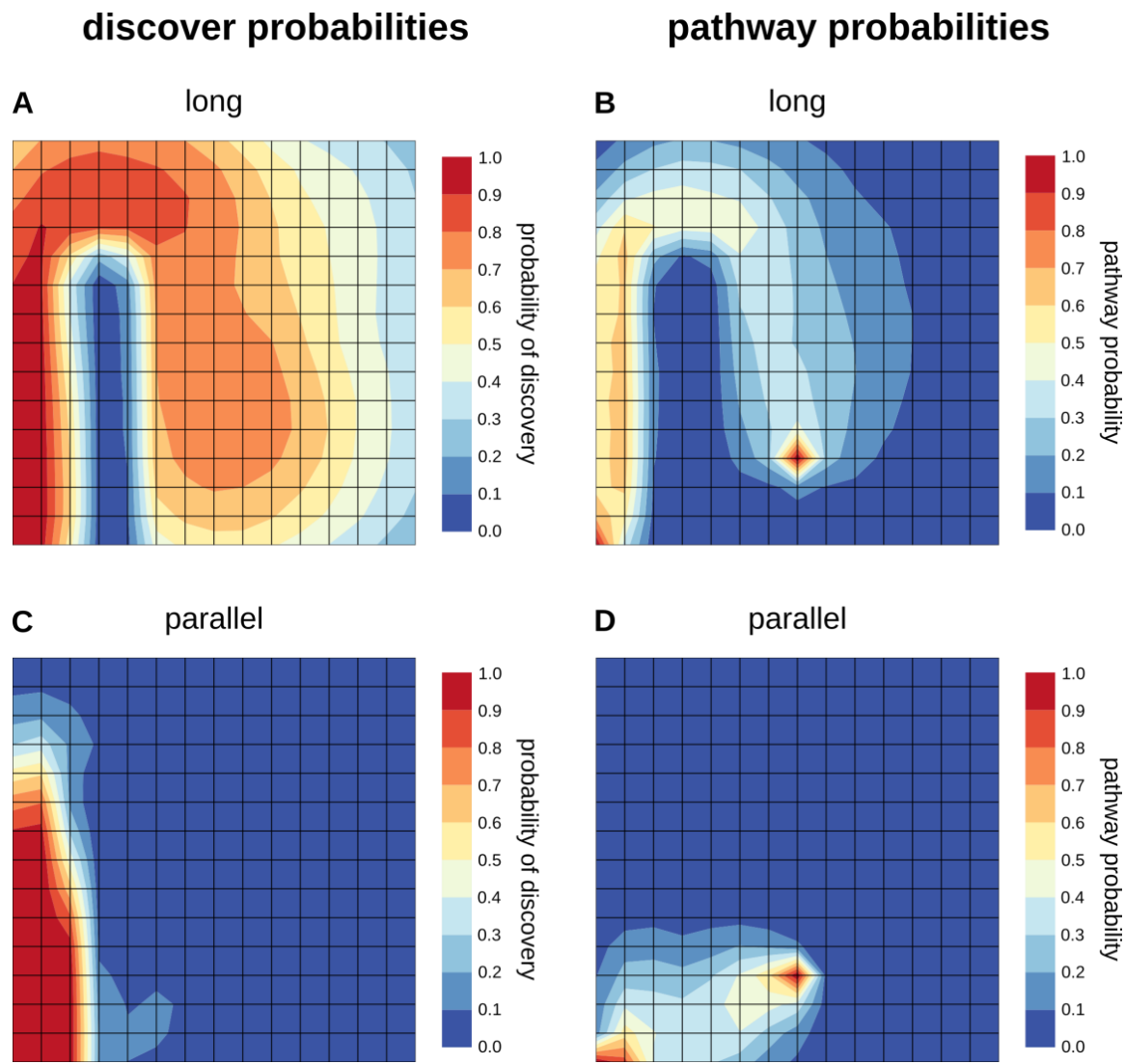

Figure 7: The discover probabilities and predicted pathways for long and parallel simulations on the landscape in Figure 6A. (A) The probability that a long simulation discovers a particular state. (B) The probability that a long simulation will predict a state to be in the highest-flux path from the start to the target state. (C) The probability that a set of parallel simulations discovers a particular state. (D) The probability that a set of parallel simulations will predict a state to be in the highest-flux path from the start to the target state.

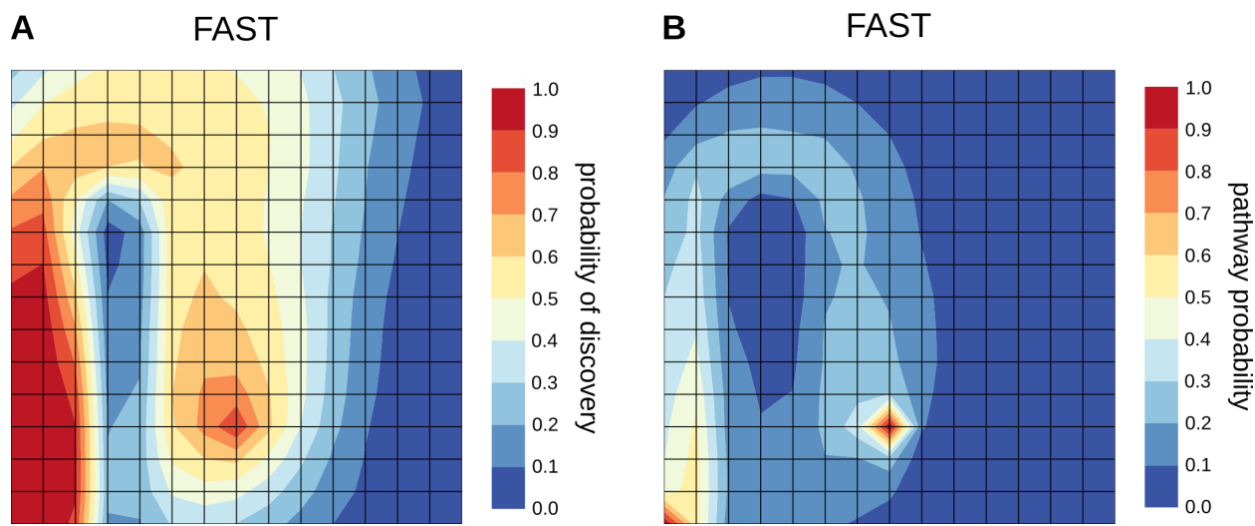

Figure 8: FAST simulations navigating a large energy barrier. (A) The probability that a FAST simulation set discovers a particular state. (B) The probability that a FAST simulation set will predict a state to be in the highest-flux path from the start to the target state.

To explore the utility of FAST when the preferred pathway is suboptimally described by the geometric ranking function, we modeled a landscape with a large barrier separating the start and target states (Figure 6). Here, the three highest-flux pathways all circumnavigate this large barrier rather than taking the geometrically shortest path (across the barrier). Indeed, the long simulations also indicate that the preferred pathway does not cut across the barrier, but follows the longer, low-energy route (Figure 7A-B).

This landscape highlights a potential pathology of running many short parallel simulations, which consistently predict that the highest-flux pathway cuts across the high-energy barrier. From Figure 7C, we observe that the probability that one of the short simulations completes the long path is significantly less than the probability that it hops across the high energy barrier. This leads to the prediction of a very unrealistic highest-flux pathway, as shown in Figure 7D. We name this undesired phenomenon "pathway tunneling", due to its loose similarity to the tunneling through high-energy barriers observed in quantum mechanics. If the length of all the parallel simulations is increased, the probability of pathway tunneling falls monotonically and converges on the correct mechanism.

From the discover probabilities in Figure 8A, we observe that FAST has a significantly higher probability of discovering the states along the preferred pathway compared to those of the tunneled pathway. It appears that even in the extreme case where the directed component is at times orthogonal to the preferred pathway, FAST's statistical component mitigates pathway tunneling. This is evidenced from counts-based adaptive samplings' ability to discover the correct pathway (Figure S6-7). However, despite this benefit, compared to the long simulations there is an increased probability of discovering the tunneled states. This isn't an issue if the estimates of the transition probabilities are accurate enough to distinguish the likelihood of each path, although the pathway probabilities in Figure 8B show that FAST non-negligibly predicts the tunneled pathway as the preferred pathway.

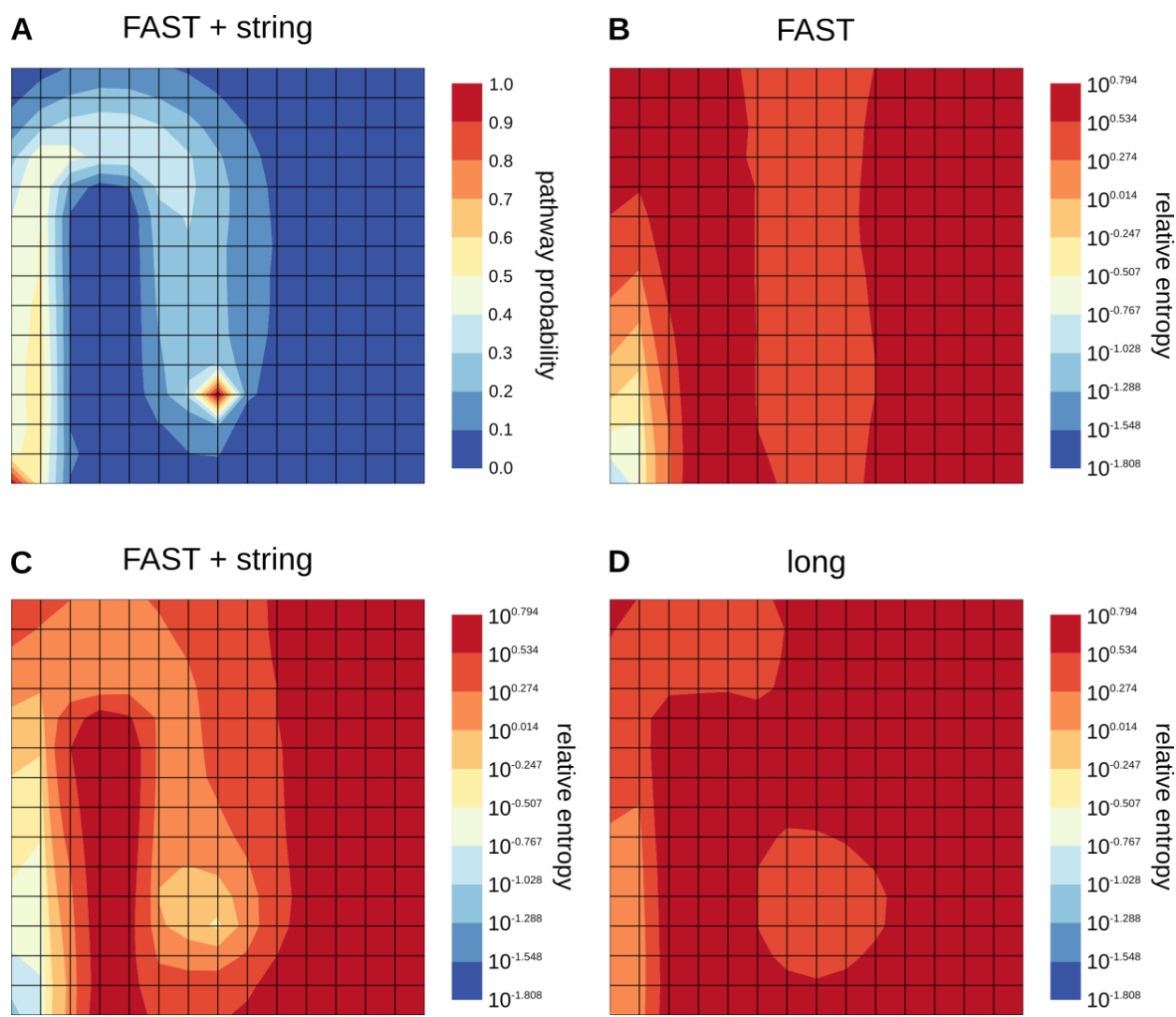

Figure 9: A comparison of predicted pathways and estimated transition probabilities between sampling methods on a landscape with a large barrier. (A) The probability that a FAST-string simulation set will predict a state to be in the highest-flux path from the start to the target state. (B-D) The Kullbeck-Liebler divergence of each states conditional transition probabilities to the true transition probabilities. Here, a lower value indicates a lower deviation from the true underlying landscape. Compared are FAST simulations, FAST simulations followed by FAST-string, and a long simulation. Each of these are produced from equivalent aggregate simulation.

*FAST-string quickly discriminates between alternative pathways*

To minimize the probability that FAST falls victim to pathway tunneling, we introduce a new ranking scheme for FAST that refines the transition probabilities along the highest-flux pathways to quantify their relative weights. This method draws inspiration from the string method,[45-47] which refines a proposed transition path by iteratively running short molecular dynamics simulations from regularly spaced conformations along the path and letting them relax towards the true lowest free energy path. Here, we begin FAST-string after first discovering a pathway, or set of pathways, to the target state using the original FAST rankings. Then, we change the ranking function to focus on refining the transition probabilities of the path(s) found. Specifically, we calculate the *n*-highest-flux pathways and rank states found in these paths by some statistical criterion. Thus, our state rankings become:

$$r(i) = \begin{cases} \bar{\psi}(i) & if\ i \in \{w_0, \ldots, w_n\} \\ 0 & otherwise \end{cases} \quad [7]$$

where $r(i)$ is the ranking of state $i$, $\bar{\psi}(i)$ is the scaled statistical component of the original FAST ranking function, and $\{w_0, \ldots, w_n\}$ represents the states found in the *n*-highest-flux paths. For our purposes, we use the counts of each state as our statistical component to favor less explored regions of the predicted pathways. We expect that sampling along these states will distinguish favorable paths from unfavorable, if multiple paths are discovered, and help relax the pathway to the preferred path if pathway tunneling has occurred.

With our FAST-string method, we are able to consistently determine the preferred transition path. Figure 9A shows that the tunneled pathway is no longer predicted as the transition pathway. We obtain this result with the same amount of aggregate simulation as the other methods; we run FAST until it discovers the end state, then switch to FAST-string for the remainder of the rounds. Instead of redundantly sampling around the target state once found, FAST-string productively refines estimates of the transition probabilities. From Figure 9B-D, we can see that FAST-string has the most accurate estimates of transition probabilities along the highest-flux pathways.

*Normalizing raw counts is the best estimator for building MSMs from adaptive sampling data*
In addition to comparing different sampling methods, it is important to ask what the best way of estimating the transition probabilities between states from a given data set is. In other words, what is the best way to use a count-matrix, which counts the observed transitions between every pair of states, to estimate the transition probabilities and equilibrium populations of each state?

The simplest way is to normalize each row in the count-matrix to get an unbiased estimate of each states conditional transition probabilities, where the first eigenvector provides the equilibrium populations.[48] However, this approach does not guarantee microscopic reversibility and can have serious pathologies if the transition probability matrix is not ergodic, especially if transitions are observed from state $n_i$ to $n_j$ but not in the opposite direction. To alleviate this issue, it is customary to assume that, prior to observing any data, each state has equal probability to transition to any other state. This can be represented by adding a pseudo-count, $\tilde{C}$, to each possible transition,

$$T_{ij}^{normalize} = \frac{C_{ij} + \tilde{C}}{\sum_k (C_{ik} + \tilde{C})} \quad [8]$$

where,

$$\tilde{C} = \frac{1}{n} \quad [9]$$

and $n$ is the number of states. An alternative estimator, called the transpose method, enforces detailed balance. At equilibrium, we know that each state transition should be equally populated by the reverse process (running an infinitely long simulation in reverse should not alter the estimates for transition probabilities). Enforcing this is straightforward, by averaging with the transpose of the count-matrix:

$$C_{ij}^{transpose} = \frac{C_{ij}+C_{ji}}{2}$$

and

$$T_{ij}^{transpose} = \frac{C_{ij}^{transpose}}{\sum_k C_{ik}^{transpose}}$$

and the equilibrium populations are calculated as,

$$\pi_i = \frac{\sum_j C_{ij}^{transpose}}{\sum_{k,j} C_{kj}^{transpose}} \quad [10]$$

More sophisticated methods have also been developed to enforce detailed balance, such as the use of maximum likelihood estimation (MLE),[21,24] and the observable operator model (OOM).[49] In the MLE method, the likelihood of the transition probability matrix given an observed trajectory, **X**, is determined to be,

$$P(T|\mathbf{X}) \propto \prod_{i,j} T_{ij}^{C_{ij}}$$

Consequently, the most likely transition probability matrix is solved as,

$$T_{ij}^{MLE} = \arg\max_{T_{ij}^*} P(T_{ij}^*|\mathbf{X})$$

A variant of the MLE method, which we will refer to as MLE-CP (MLE- with Constrained Populations), has also been developed to enforce a pre-determined equilibrium probability distribution. This is useful with experimental estimates of state populations.[50] Lastly, the OOM was recently developed as a generalization to hidden Markov models,[51] and restructured for use with MSMs.[49,52]

Each of these methods has been studied theoretically, in the limit of infinite data, and on small systems where sampling is not an issue. However, we are interested in the likely scenario where sampling is far from exhaustive. To test MSM construction in this regime, we used the FAST simulation sets on the landscape in Figure 5A to generate MSMs using the five methods listed above: 1) normalize, 2) transpose, 3) MLE, 4) OOM, and 5) MLE with constrained populations. We then compared the MSMs predictions of thermodynamics (equilibrium populations) and kinetics (transition probabilities) to the true distributions calculated from the underlying landscape.

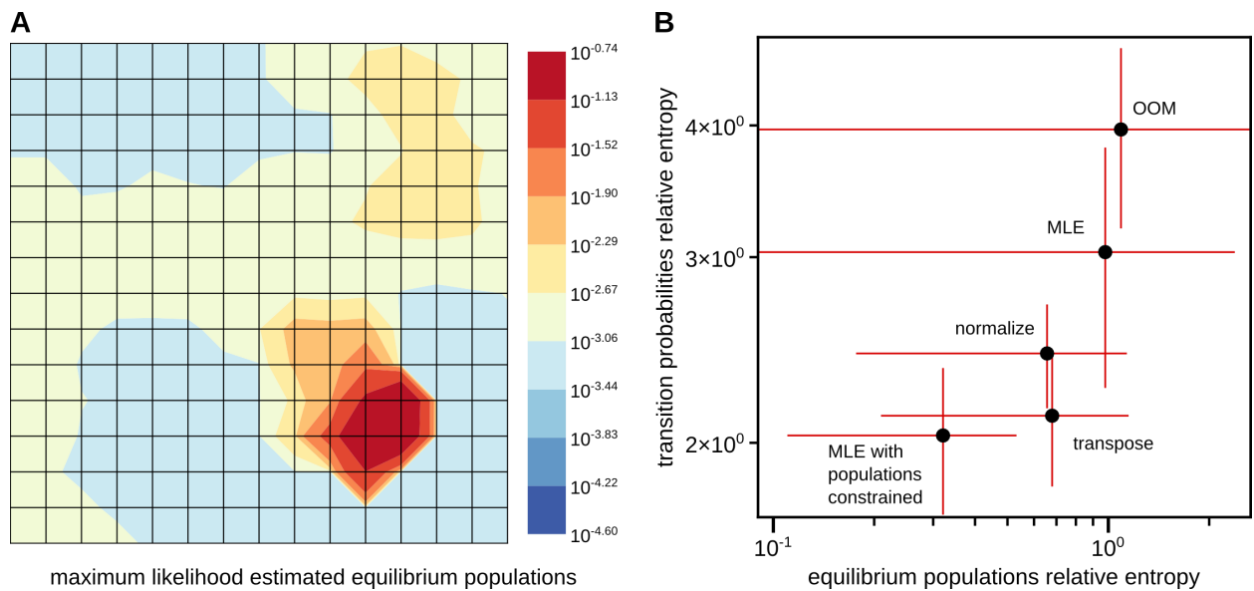

Figure 10: An analysis of MSM estimators' performance on the landscape depicted in Figure 5A. (A) The predicted state populations for a single FAST simulation using the MLE method. (B) A comparison of the MLE-CP, transpose, normalize, MLE, and OOM estimators. Solid points are the average relative entropy for transition probabilities and equilibrium populations. Red lines are the standard deviations of these values.

Upon inspection of the predicted equilibrium populations, we find that the MLE and OOM methods have a tendency to significantly over predict the populations of an arbitrary set of states. Figure 10A is an example of this phenomenon for MLE, where four states are predicted to have a total probability of 0.58 even though the probability that they were sampled in the raw data is only 0.016. For reference, the true total probability of these states is 0.029 and the true probability of the most populated state in the underlying landscape is 0.032. In comparison, Figure S9 shows that normalize and transpose give more reasonable predictions. Characterizing this over the entire dataset, we observe that on average, the largest predicted state population for MLE and OOM is 10.5 ± 21.0 and 16.4 ± 53.5 times larger than its true population. Interestingly, the deviation in these predictions are sizable; the most egregious observances of an overinflated state population for MLE and OOM were predictions of a single state containing 0.38 and 0.55 of the total population for each method, respectively. Additionally, MLE and OOM do not regularly overpopulate the same state; the probability that the state with the largest predicted population is truly the most populated state is 0.16 and 0.14 for MLE and OOM, respectively, compared to 0.35 and 0.33 for normalize and transpose, respectively. On the other hand, normalize and transpose have a largest populated state that is only 2.2 ± 2.5 and 2.0 ± 2.7 times its true population. To see how this behavior affects predictions for all states, we compute the relative entropy between each models' prediction of transition probabilities and equilibrium populations to the true distributions.

The MLE-CP is shown to generate an MSM with the most accurate estimates of kinetics and thermodynamics for FAST simulations on this particular landscape. Figure 10B shows the average deviations of transition probabilities and equilibrium populations from the true values for the underlying landscape for each MSM method. It is not surprising that constraining the populations to their true values performs well. More surprisingly, though, are the significant improvements to the transition probabilities when the equilibrium populations are constrained.

However, *a priori* knowledge of the equilibrium distribution is not typically available, so it is not currently possible to adopt this approach as standard practice.

The normalize and transpose methods produce the next most accurate estimates of transition probabilities and equilibrium populations. However, despite transposes' adequate performance on this landscape, it can be shown from equation 10 that the estimated equilibrium populations are directly related to the amount of sampling in each state. This is not thought to be ideal with adaptive sampling, since continually sampling from a state will artificially inflate its estimated equilibrium population. Transpose does well on this particular landscape due to the relatively flat energy surface of the preferred path and would be less favorable with real landscapes. Therefore, we recommend the use of the normalize method with adaptive sampling data for its simplicity and accurate estimates of thermodynamics and kinetics.

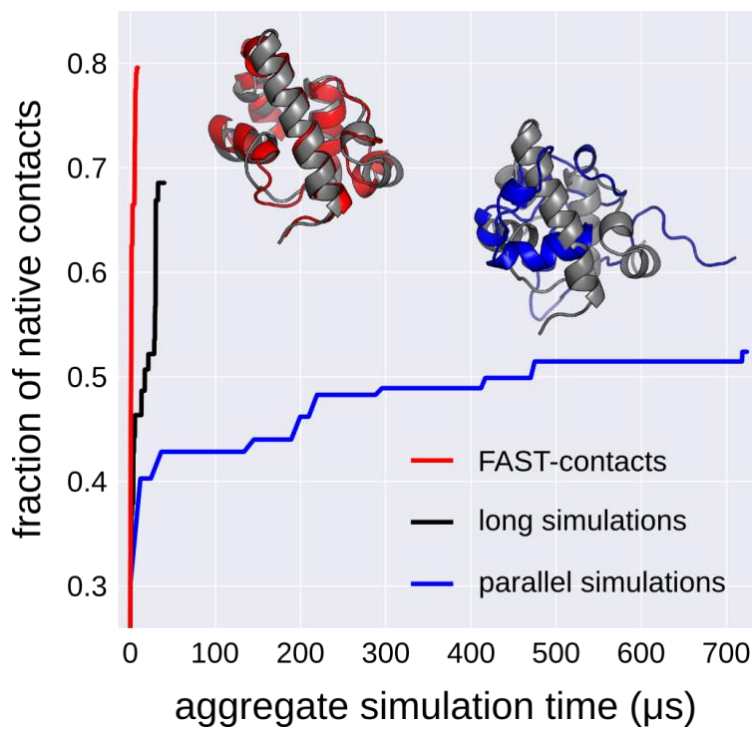

Figure 11: The largest observed fraction of native contacts as a function of aggregate simulation time for three equilibrium-based sampling methods. Simulation sets were generated from the same initial structure, which had a fraction of native contacts of 0.17 formed. Structures indicate the largest fraction of native contacts observed in a single run of FAST (red) or parallel simulations (blue) in contrast with the crystal structure (gray) (PDBID: 1LMB).

*Simulations of λ-repressor recapitulate the patterns observed for simple landscapes*
Kinetic Monte Carlo simulations on physically inspired landscapes have provided valuable functional insight, but it is important to ensure that our conclusions hold true for the exploration of real protein landscapes. Protein conformational landscapes are hyper-dimensional and likely have many barriers, both enthalpic and entropic. Thus, we turn to using all-atom MD simulations for three sampling methods: 1) long simulations, 2) massively parallel simulations, and 3) FAST-contacts (which ranks states by the fraction of native contacts that are present). Each method uses the same unfolded starting structure and simulation parameters, where extended details are described in the Methods. As for a model system, we chose to simulate a fast-folding variant of the λ-repressor.[53] Due to its speed of folding and size, the kinetics of this protein have been extensively studied, both experimentally and computationally, making it ideal for use when comparing sampling strategies.

Unlike the simple landscapes in previous sections, all-atom MD simulations are computationally expensive and sample along vast conformational landscapes. As a consequence, we cannot run thousands of iterations to robustly characterize the probability of discovering a particular state. Instead, we can compare the performance of each method by focusing on a more coarse-grained metric of interest, such as the computational time required to reach the folded state, as measured by the fraction of native contacts present.

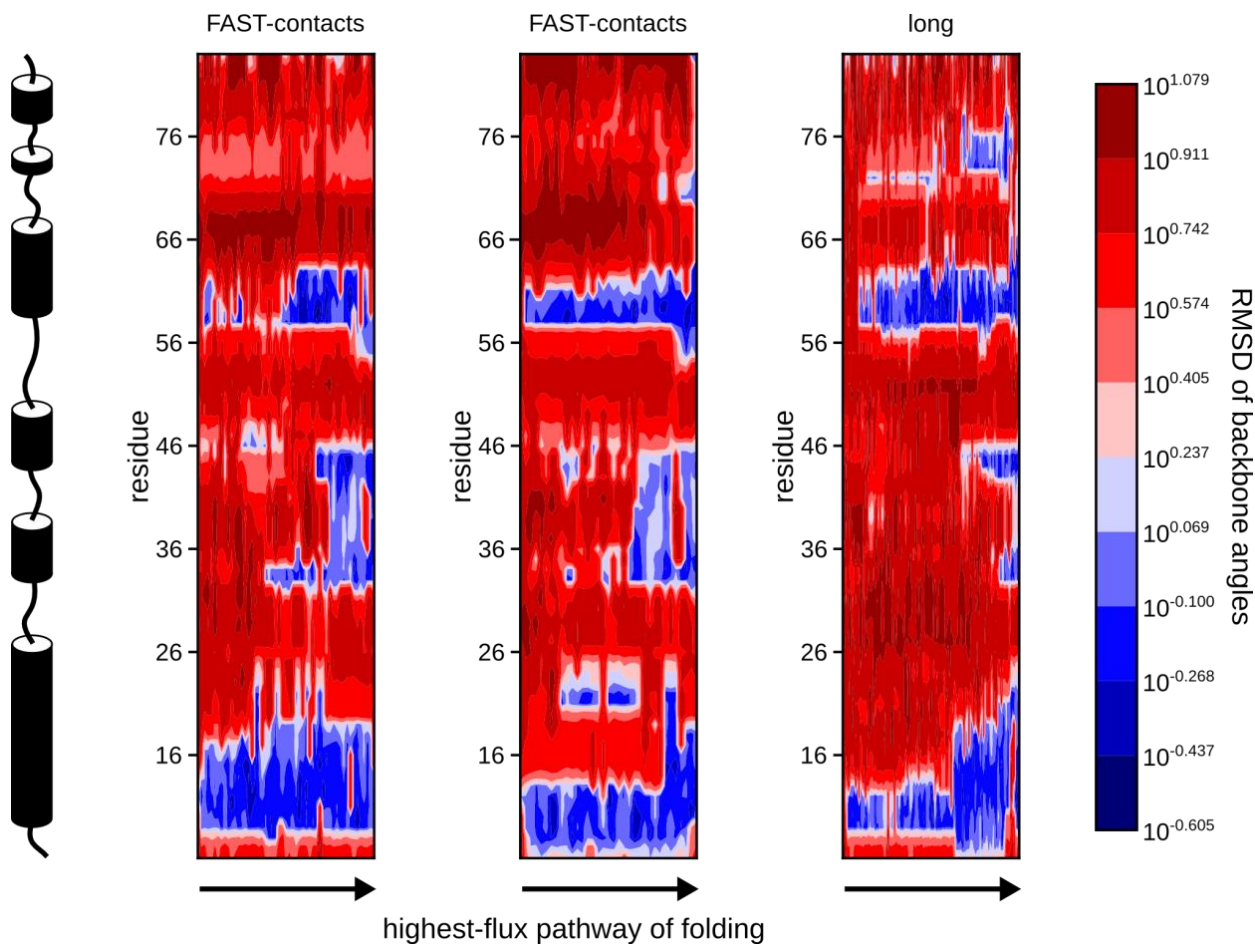

Figure 12: Analysis of predicted folding pathways for λ-repressor using the RMSD of each residues' backbone $\phi$ and $\psi$ angles to the crystal structure (PDBID: 1LMB). Folding pathways are defined as an MSMs' highest-flux path from the starting state to the state with the largest fraction of native contacts. The time evolution of each residue's backbone RMSDs are shown along the x-axis for the predicted folding pathway from two separate runs of FAST-contacts and a single set of long simulations.

   Analysis of the three sampling methods reveals that adaptive sampling yields similar benefits to those found on our simple landscapes. Figure 11 shows the highest fraction of native contacts observed for each sampling method as a function of the aggregate simulation time. Remarkably, FAST-contacts folds the λ-repressor with ~4 μs of aggregate simulation, which is faster than its experimental folding time. By comparison, it takes nearly 40 μs of long simulations to achieve a similar level of foldedness. Furthermore, the massively parallelized simulations, with over 700 μs of aggregate simulation time, do not discover the folded state. These results are in strong agreement with the discovery predictions from the landscapes in Figures 3 and 5.

In addition to understanding the probability of observing a folded state, we are interested in the predicted folding pathways. However, the idea of characterizing a pathway for all-atom MD simulations is more complicated than on the theoretical landscapes; state-space is significantly larger, computational limitations prevent multiple trials to assess the stochasticity, and the optimal (human-intuitive) parameters to define a pathway are not straightforward. The long and parallel simulations require too much computational resources to gather statistics on, although we were able to generate five independent trials of FAST in a reasonable timeframe. For the purposes of defining a pathway, others have successfully taken the approach of characterizing folding by the rate of formation of secondary structural elements.[54-56] Thus, we also aim to characterize the rate of secondary structure formation by determining each residues' root-mean-square deviation (RMSD) of backbone dihedrals from the crystal structure, for states along the predicted highest-flux pathway. We plot these deviations for two representative runs of FAST-contacts and the long simulations in Figure 12.

The predicted pathways for each of these methods are reasonably consistent with one another. FAST-contacts predominantly predicts the folding of helices 1 and 4 before helices 2 and 3. This is consistent with our prediction using the single set of long simulations. Additionally, this is what has been seen with previous simulation reports,[57,58] and hydrogen exchange experiments.[59] Interestingly, this is counter to the results from a Gō model, which has been previously used and describes helices 1-4 folding cooperatively.[60] This difference suggests that FAST-contacts is not simply an expensive Gō model.

**Conclusions**

We have presented a systematic comparison of different sampling strategies on a variety of representative energy landscapes. We first developed an analytic expression for the probability of discovering states on a landscape that depends on the number, length, and starting state of simulations. From this we find that long simulations have a higher probability of discovering states on landscapes with reduced dimensionality, though parallel simulations have a higher probability of discovering states as the dimensionality increases. To build upon this, we used kinetic Monte Carlo simulations on more complex landscapes to compare four sampling strategies (long simulations, parallel simulations, counts adaptive sampling, and FAST), which each reveal a unique state discovery signature. Understanding the differences in how these sampling strategies discover states has provided insight into their advantages and disadvantages. Specifically, long simulations provide an unbiased estimates of transition paths, although requires significant computational resources compared to adaptive sampling or FAST and produces less accurate MSMs. Parallel simulations thoroughly explore around the starting state and provide excellent estimates of transition probabilities (for the states discovered) but are unlikely to explore distant regions of conformational space and may provide erroneous transition paths. Counts-based adaptive sampling discovers the most states along a variety of paths, although these states are likely to be unproductive for a given goal, especially on landscapes with large dimensionality.

Throughout our analysis, we have taken special interest in the performance of our recently developed goal-oriented sampling algorithm, FAST. On our simple landscapes, we find that FAST consistently has the highest probability of discovering a target state, predicts reasonable pathways, and provides the best estimates of transition probabilities for an entire MSM as well as of the true highest-flux pathway (Table S1). Furthermore, we demonstrate the utility of FAST using all-atom MD simulations of the λ-repressor. FAST produces an accurate folding pathway with an order of magnitude less aggregate simulation than long simulations, and orders of magnitude less than parallel simulations.

**Methods**
*Generation and simulation of simple landscapes*
The three physically inspired potential energy landscapes were generated by selectively adding Gaussian potentials to an otherwise flat surface. These potential energy landscapes were then converted to a transition probability matrix using the following relations:

$$\zeta_{ij} = \begin{cases} e^{\varepsilon_i - \varepsilon_j} & if\ \varepsilon_i < \varepsilon_j \\ 1 & if\ \varepsilon_i \geq \varepsilon_j \end{cases}$$

for all $j$ that are neighbors of $i$, and where $\varepsilon_i$ is the potential energy of state $n_i$ in units of $k_B T$. This can then be row-normalized to obtain,

$$T_{ij} = \frac{\zeta_{ij}}{\sum_j \zeta_{ij}}$$

Kinetic Monte Carlo simulations were then performed with this transition probability matrix for four sampling schemes: 1) long simulations, 2) parallel simulations, 3) counts-based adaptive sampling, and 4) FAST simulations. For each of the sampling schemes, 5,000 independent sets of simulations were generated, each with a total of 1,000 time-steps. For the long simulations, this consisted of 5,000 single trajectories, of 1,000 steps. Each set for the parallel simulations consisted of 25 trajectories with 40 steps.

Counts-based adaptive sampling and FAST both followed the same basic protocol: 1) generate 5 trajectories of 20 steps each from the initial state, 2) build an MSM, 3) rank states, 4) generate 5 more trajectories of 20 steps each from the top 5 states with the highest ranking, 5) repeat steps 2-4 for a total of 10 rounds. The difference between counts-based adaptive sampling and FAST is in the manner of ranking states between each round. For counts adaptive sampling, states were ranked by their observed counts in the MSM, with lower counts being more favorable. For FAST, we used the following ranking,

$$r_\phi(i) = \bar{\phi}(i) + \alpha\bar{\psi}(i) + \beta\chi(i) \qquad [11]$$

where $\bar{\phi}$ is the feature-scaled directed component (Euclidean distance to the target state), $\bar{\psi}$ is the feature scaled undirected component, $\chi$ is a similarity penalty, and $\alpha$ and $\beta$ control the weights of $\bar{\psi}$ and $\chi$, respectively, as has been published previously.[19] Here, $\bar{\psi}(i)$ is taken to be the state counts and a value of 1 was used for both $\alpha$ and $\beta$. The directed component for each state on the landscapes was the grid distance to the target state. The similarity penalty for each state selected is defined with,

$$\chi(i) = \begin{cases} 0 & if\ N = 0 \\ \frac{1}{N}\sum_{j=1}^{N}\left(1 - e^{\frac{-d_{ij}^2}{2w^2}}\right) & if\ N > 0 \end{cases} \quad [12]$$

which is the average of the Gaussian weighted grid distance, $d$, from state $n_i$ to the $N$ states that have been selected for reseeding so far, where $w$ is the Gaussian width (set to the clustering radius). Thus, selecting states proceeds as follows: 1) rank all states by the FAST ranking and select the top state, 2) add the similarity penalty and select the top-ranking state as the next state, 3) repeat step 2 until the desired number of states have been selected.

After generating the state trajectories on the landscapes from the sampling methods, state discover probabilities, pathway probabilities, and relative entropies were calculated. The discover probabilities were calculated by first using equation 1 to indicate if a state was discovered for each simulation set. These values for $D_{ij}^{K,M}$ were then averaged over the 5,000 trials to determine the probability of discovering a state in the simulation set, $P(D_{ij}^{K,M} = 1)$. Similar to the discover probabilities, the pathway probabilities were calculated by averaging the output of a selector function, over the simulation sets, that indicated if a state was present in the predicted highest-flux pathway. The highest-flux pathway for each simulation set was calculated using MSMBuilder.[61] The relative entropies of each state were calculated as the Kullback-Leibler divergence between the estimated conditional transition probabilities from that state and those of the underlying energy distribution:

$$D_{KL}^i(P_i||Q_i) = -\sum_i P_i \log\left(\frac{Q_i}{P_i}\right)$$

where $D_{KL}^i$ is the relative entropy for state $i$, $P_i$ is the $i$-th row of the true transition probability matrix, and $Q_i$ is the $i$-th row of the transition probability matrix reconstructed from synthetic trajectories. The relative entropy of the entire MSM is a population weighted average of these values, as is described previously.[33,36] MSMs were constructed with either the MSMBuilder or PyEMMA software packages.[61-63]

*Molecular Dynamics Simulations*
Three sets of all-atom molecular dynamics simulations for the λ-repressor were generated: 1) 7,005 parallel simulations (103.4 ± 82.0 ns each), 2) 16 long simulations (2.5 μs each), and 3) FAST-contacts simulations (30 rounds of 10 simulations per round, with 30 ns per simulation). Each of these simulations were run with Gromacs 5.1.1[64] using the AMBER03 force field with explicit TIP3P solvent.[65,66]

Each of these sets of simulations began from the same starting structure, which was prepared as follows. First, a linear structure of the λD14A mutant[53] was generated using the VMD software package.[67] The linear structure was equilibrated for 1 ns at 420 K with OBC GBSA implicit solvent.[68] The final conformation was then placed in a dodecahedron box that extended 1.0 nm beyond the protein in any dimension. This system was then energy minimized with the steepest descent algorithm until the maximum force fell below 100 kJ/mol/nm using a step size of 0.01 nm and a cutoff distance of 1.2 nm for the neighbor list, Coulomb interactions, and van der Waals interactions.

For production runs, all bonds were constrained with the LINCS algorithm and virtual sites were used to allow a 4 fs time step. Cutoffs of 1.0 nm were used for the neighbor list, Coloumb interactions, and van der Waals interactions. The Verlet cutoff scheme was used for the neighbor list. The stochastic velocity rescaling ($v$-rescale) thermostat was used to hold the temperature at 360 K and conformations were stored every 50 ps.[69]

*FAST Simulations*

Five sets of FAST-contacts simulations were generated that each observed an independent folding trajectory for the λ-repressor. Each set of FAST-contacts consisted of 9 μs of aggregate simulation time: 30 rounds, of 10 simulations per round, where each simulation was 30 ns. Between each round, discrete states were generated by clustering atomic coordinates of backbone atoms using a $k$-centers algorithm based on RMSD between conformations until every cluster center had a radius less than 3.0 Å. States were selected for reseeding based on the ranking function and selection criterion described with equations 11 and 12. The similarity penalty used was RMSD between cluster centers, where the Gaussian width, $w$, was set to the clustering radius of 3.0 Å. The directed component to the FAST ranking was the feature scaled values of the fraction of native contacts, described elsewhere.[70]

*MSM Construction and Analysis*

MSMs were built of each simulation set using MSMBuilder.[61,62] The construction of each MSM followed the same basic protocol: 1) cluster conformations into discrete states, 2) count transitions between these states at a specified lag-time, and 3) generate each states' conditional transition probabilities. For the first step, atomic coordinates of backbone heavy atoms (CO, $C_\alpha$, O, N) and $C_\beta$ atoms were clustered with a $k$-centers clustering algorithm until every cluster center had a radius of less than 3.0 Å. A lag-time of 5 ns was used for counting transitions between states. Each states' conditional transition probabilities were computed using the normalize method with a prior-counts, as described with equations 8 and 9. Structural analysis was aided with the use of MDTraj.[71]

**Acknowledgements**

This work was funded by NSF CAREER Award MCB-1552471 and NIH R01GM12400701. G.R.B. holds a Career Award at the Scientific Interface from the Burroughs Wellcome Fund and a Packard Fellowship for Science and Engineering from The David and Lucile Packard Foundation. M.I.Z. holds a Monsanto Graduate Fellowship and a Center for Biological Systems Engineering Fellowship.

**Supporting information**
*Calculating discover probabilities example*

Given the landscape depicted in Figure 1B, with the transition probability matrix,

$$T_{ij} = \begin{bmatrix} 0.65 & 0.3 & 0.05 \\ 0.25 & 0.5 & 0.25 \\ 0.25 & 0.25 & 0.5 \end{bmatrix}$$

we can calculate the probability of discovering state $n_j$ from simulations starting from state $n_i$ given 3 simulations of length 2, $P\left(D_{ij}^{K=\{2,2,2\},M=3} = 1\right)$, by first calculating the probability that a single simulation of length 2 discovers state $n_j$, $P\left(D_{ij}^{K=\{2\},M=1} = 1\right)$. To do this, we use equation 4 to determine the probability of being in any of the three states at each timestep, conditional on not having discovered state $n_j$ yet. Before simulations ($k = 0$), the probability of discovering state $n_j$ is 1 if starting from state $n_j$, and 0 otherwise, which is simply the identity matrix,

$$P(v_{ij}^{k=0} = 1) = I$$

To determine the probability of being in any state after the first timestep, we propagate the probabilities with the transition probability matrix,

$$P(v_{ij}^{k=1} = 1) = P(v_{ij}^{k=0} = 1)T = T_{ij}$$

For the second step, we propagate the probabilities conditional to not having discovered state $n_j$ yet

$$P\left(v_{i'j'}^{k=2} = 1 \mid \{v_{ij}^{k'} = 0 \;\forall\; k' < 2\}\right) = P\left(v_{i'j'}^{k=1} = 1 \mid v_{ij}^{k=1} = 0\right)T$$

where, for $j = 2$, we have,

$$P\left(v_{i'j'}^{k=1} = 1 \mid v_{i2}^{k=1} = 0\right) = \begin{bmatrix} 0.68 & 0.32 & 0 \\ 0.33 & 0.67 & 0 \\ 0.5 & 0.5 & 0 \end{bmatrix}$$

which are the renormalized rows of $P(v_{ij}^{k=1} = 1)$ after setting column 2 to 0. Propagating this by the transition probability matrix, we obtain the probability of being in state 2 given that it was not discovered previously,

$$P\left(v_{i'j'}^{k=2} = 1 \mid \{v_{i2}^{k'} = 0 \;\forall\; k' < 2\}\right) = \begin{bmatrix} 0.524 & 0.363 & 0.113 \\ 0.383 & 0.433 & 0.183 \\ 0.450 & 0.400 & 0.150 \end{bmatrix}$$

Combining these probabilities of being in state 2 at various time-steps, we calculate the probability of discovering state 2, $P\left(D_{i2}^{K=\{2\},M=1} = 1\right)$, as,

$$P\left(D_{i2}^{K=\{2\},M=1} = 1\right) = 1 - \left(1 - P(v_{ij}^{k=0} = 1)\right)_{i2} * \left(1 - P(v_{ij}^{k=1} = 1)\right)_{i2} * \left(1 - \right.$$

$$\left.P\left(v_{i'j'}^{k=2} = 1 \mid \{v_{i2}^{k'} = 0 \ \forall \ k' < 2\}\right)\right)_{i2} = 1 - \begin{bmatrix} 1-0 \\ 1-0 \\ 1-1 \end{bmatrix} * \begin{bmatrix} 1-0.05 \\ 1-0.25 \\ 1-0.5 \end{bmatrix} * \begin{bmatrix} 1-0.113 \\ 1-0.183 \\ 1-0.150 \end{bmatrix} = \begin{bmatrix} 0.16 \\ 0.39 \\ 1.0 \end{bmatrix}$$

Calculating the columns for $j = \{0, 1\}$, we get the full discover probabilities between any $n_i$ and $n_j$ as,

$$P\left(D_{ij}^{K=\{2\},M=1} = 1\right) = \begin{bmatrix} 1.0 & 0.51 & 0.16 \\ 0.44 & 1.0 & 0.39 \\ 0.44 & 0.45 & 1.0 \end{bmatrix}$$

Next, this is used to calculate the discover probabilities of 3 independent simulations of length 2 using the following,

$$P(D_{ij}^{K,M} = 1) = 1 - \left[1 - P\left(D_{ij}^{K=\{2\},M=1} = 1\right)\right]^3 = 1 - \left[1 - \begin{bmatrix} 1.0 & 0.51 & 0.16 \\ 0.44 & 1.0 & 0.39 \\ 0.44 & 0.45 & 1.0 \end{bmatrix}\right]^3 =$$

$$\begin{bmatrix} 1.0 & 0.88 & 0.41 \\ 0.82 & 1.0 & 0.77 \\ 0.82 & 0.83 & 1.0 \end{bmatrix}$$

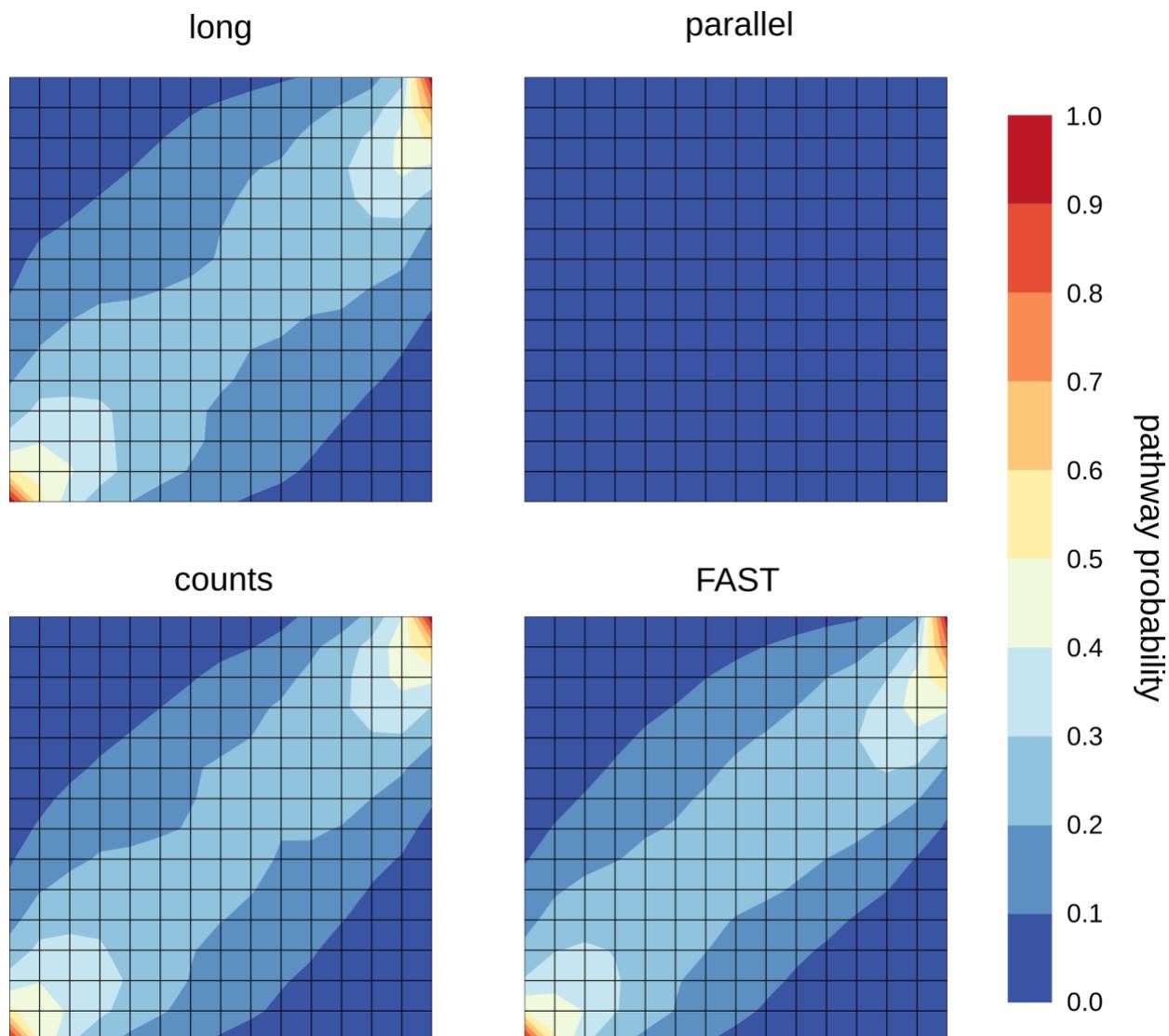

Figure S1: The pathway probabilities (the probability that a state is predicted to be in the highest-flux pathway from the start to the target) for the funneled landscape in Figure 3. Shown are the probabilities for four sampling strategies, a single long simulation, many parallel simulations, counts-based adaptive sampling, and the goal-oriented FAST simulations. The parallel simulations did not observe a transition, and thus, do not have a pathway.

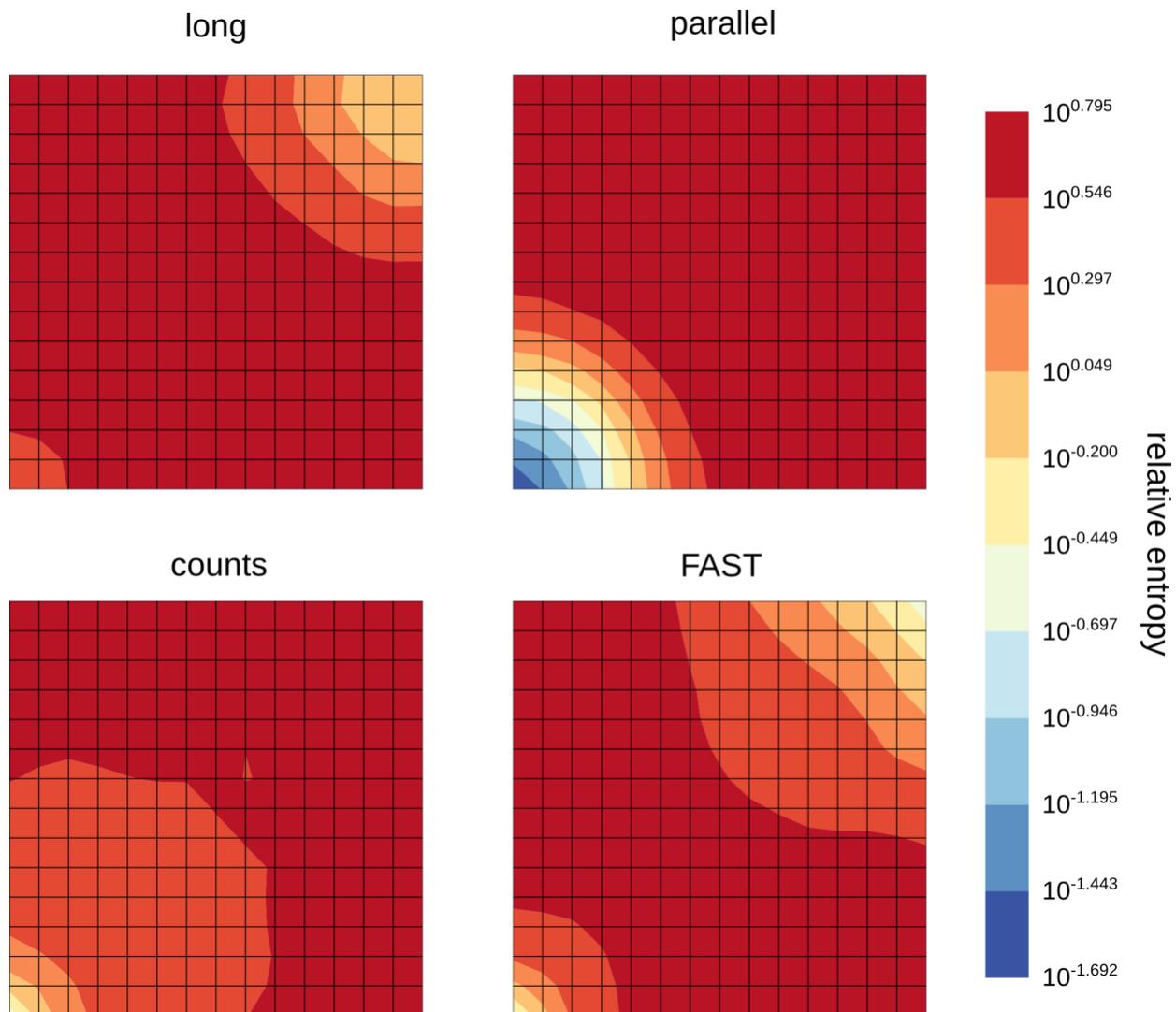

Figure S2: The average Kullbeck-Liebler divergence of each states conditional transition probabilities to the true transition probabilities for the funneled landscape in Figure 3. Shown are the average divergences of each state for four sampling strategies, a single long simulation, many parallel simulations, counts-based adaptive sampling, and the goal-oriented FAST simulations.

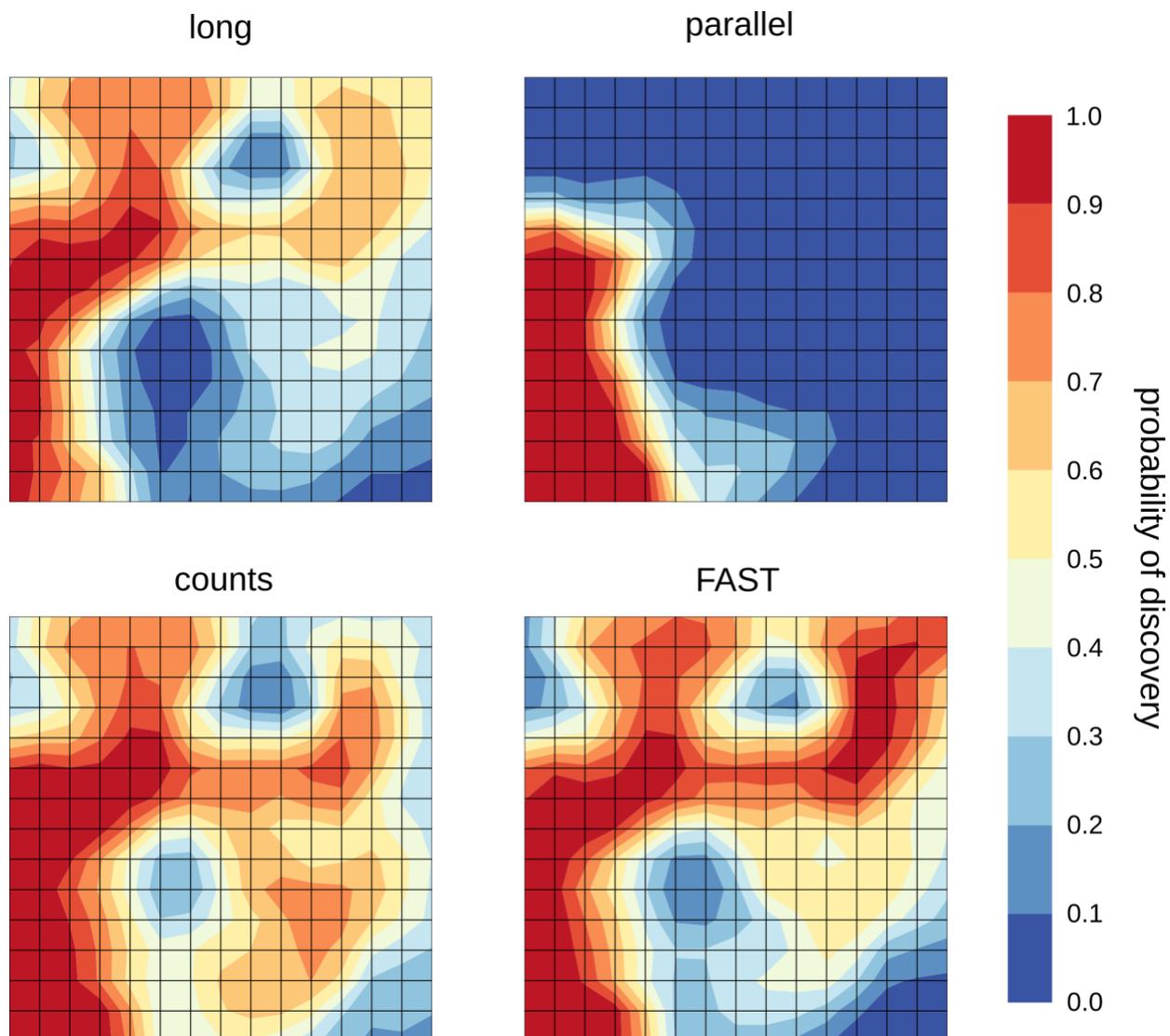

Figure S3: The discover probabilities (the probability that a simulation set observes a particular state) on the random barriered landscape in Figure 5A. Shown are the probabilities for four sampling strategies, a single long simulation, many parallel simulations, counts-based adaptive sampling, and the goal-oriented FAST simulations.

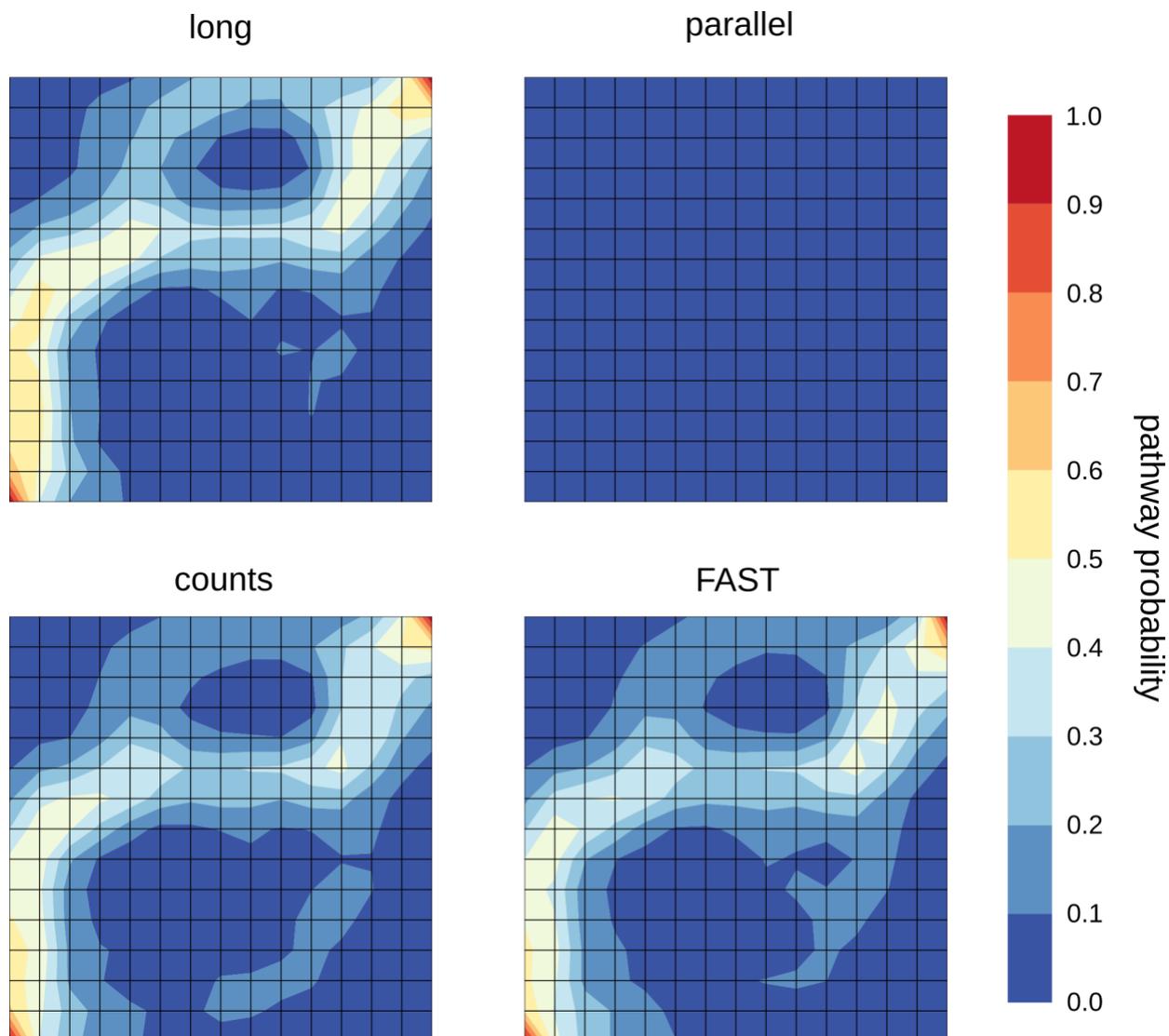

Figure S4: The pathway probabilities (the probability that a state is predicted to be in the highest-flux pathway from the start to the target) for the random barriered landscape in Figure 5A. Shown are the probabilities for four sampling strategies, a single long simulation, many parallel simulations, counts-based adaptive sampling, and the goal-oriented FAST simulations. The parallel simulations did not observe a transition, and thus, do not have a pathway.

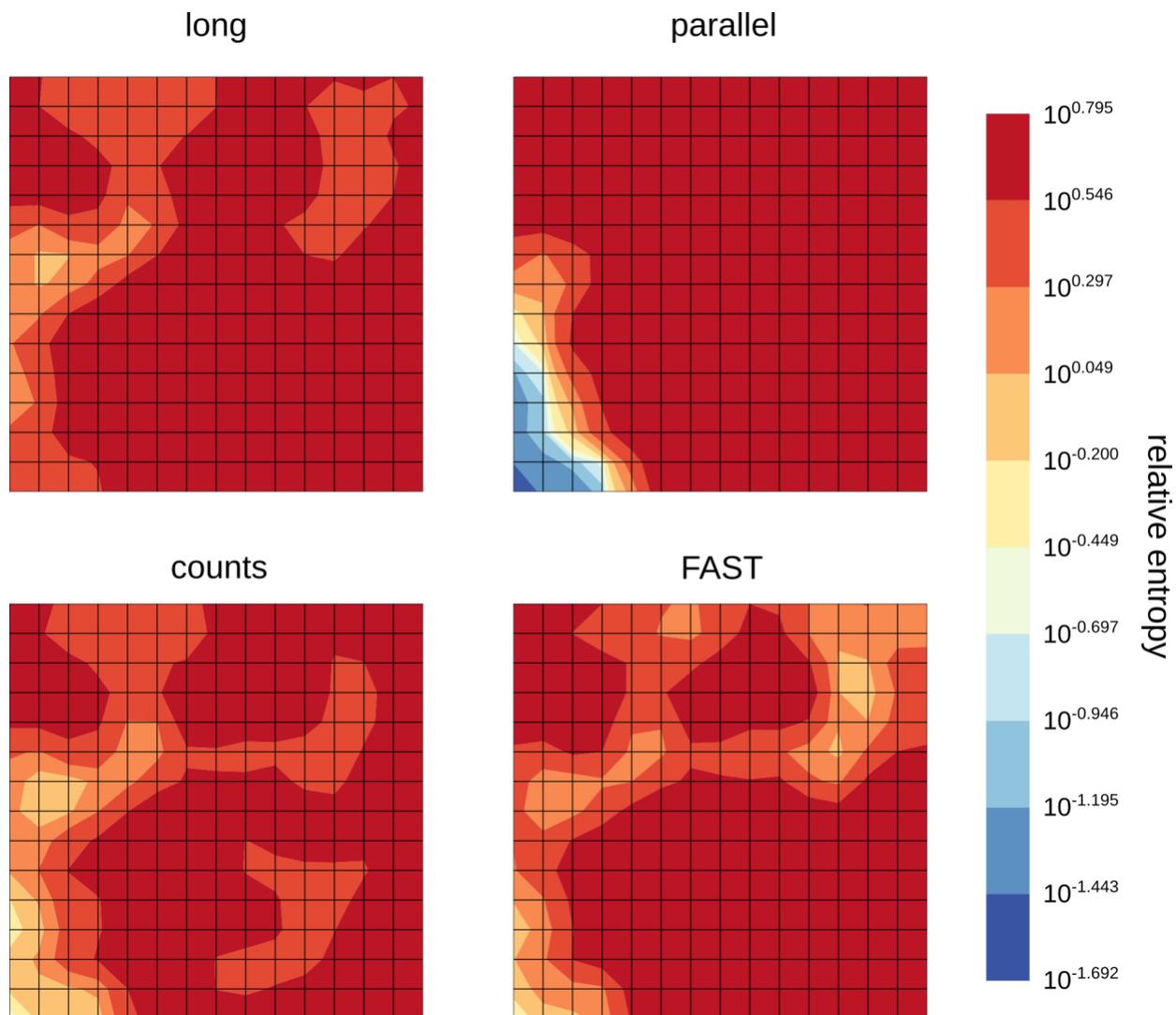

Figure S5: The average Kullbeck-Liebler divergence of each states conditional transition probabilities to the true transition probabilities for the random barriered landscape in Figure 5A. Shown are the average divergences of each state for four sampling strategies, a single long simulation, many parallel simulations, counts-based adaptive sampling, and the goal-oriented FAST simulations.

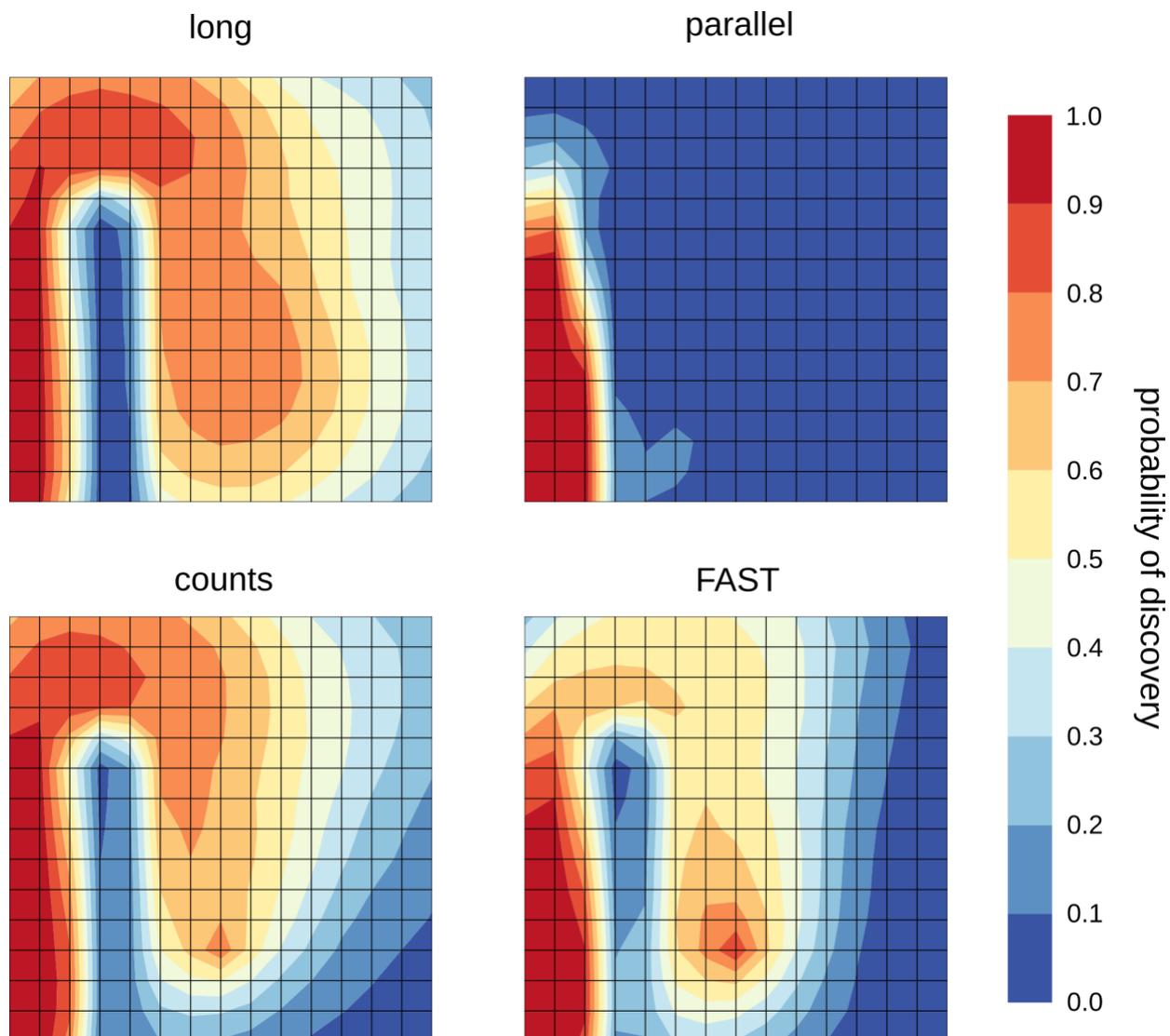

Figure S6: The discover probabilities (the probability that a simulation set observes a particular state) on the large barriered landscape in Figure 6. Shown are the probabilities for four sampling strategies, a single long simulation, many parallel simulations, counts-based adaptive sampling, and the goal-oriented FAST simulations.

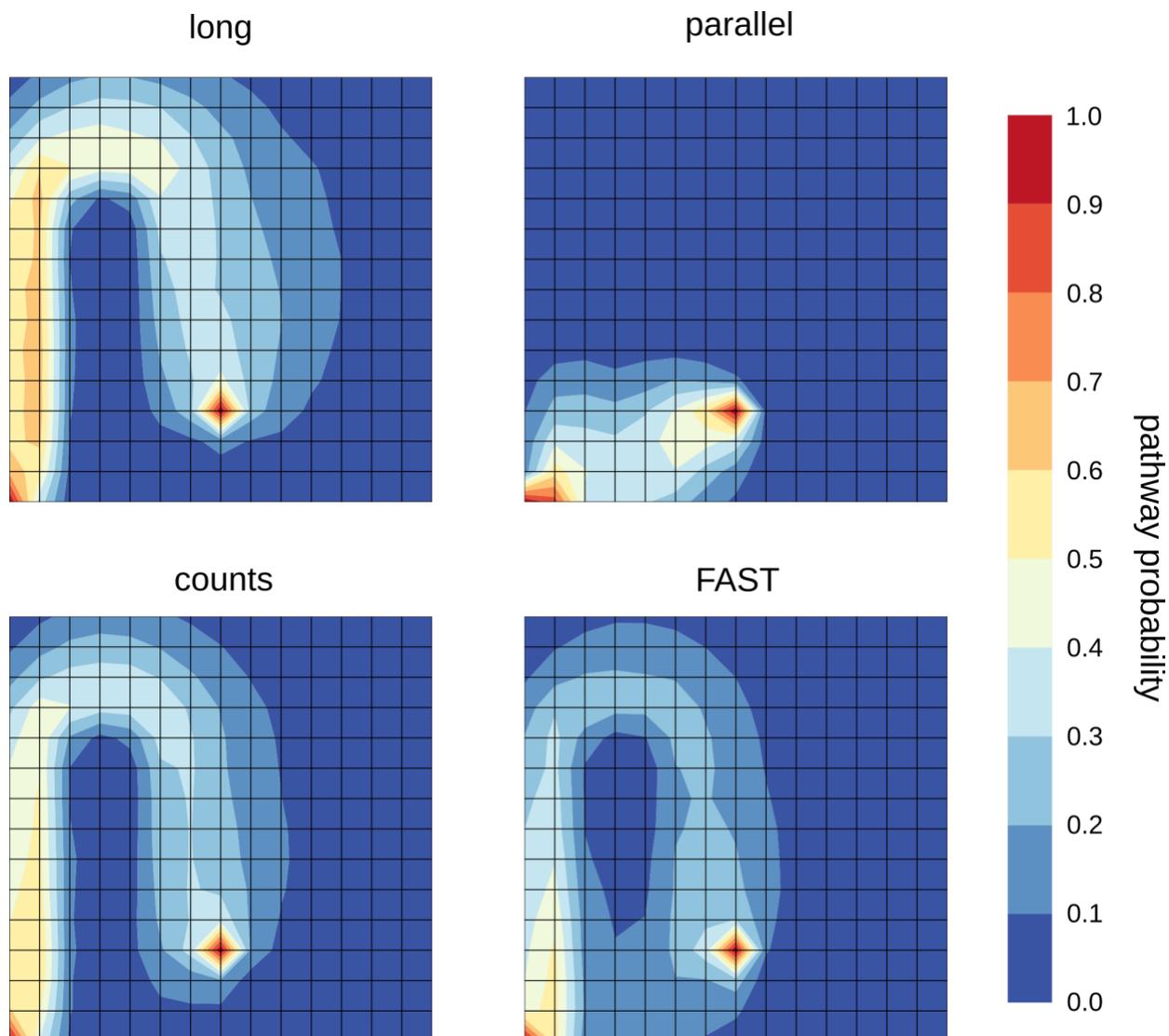

Figure S7: The pathway probabilities (the probability that a state is predicted to be in the highest-flux pathway from the start to the target) for the large barriered landscape in Figure 6. Shown are the probabilities for four sampling strategies, a single long simulation, many parallel simulations, counts-based adaptive sampling, and the goal-oriented FAST simulations.

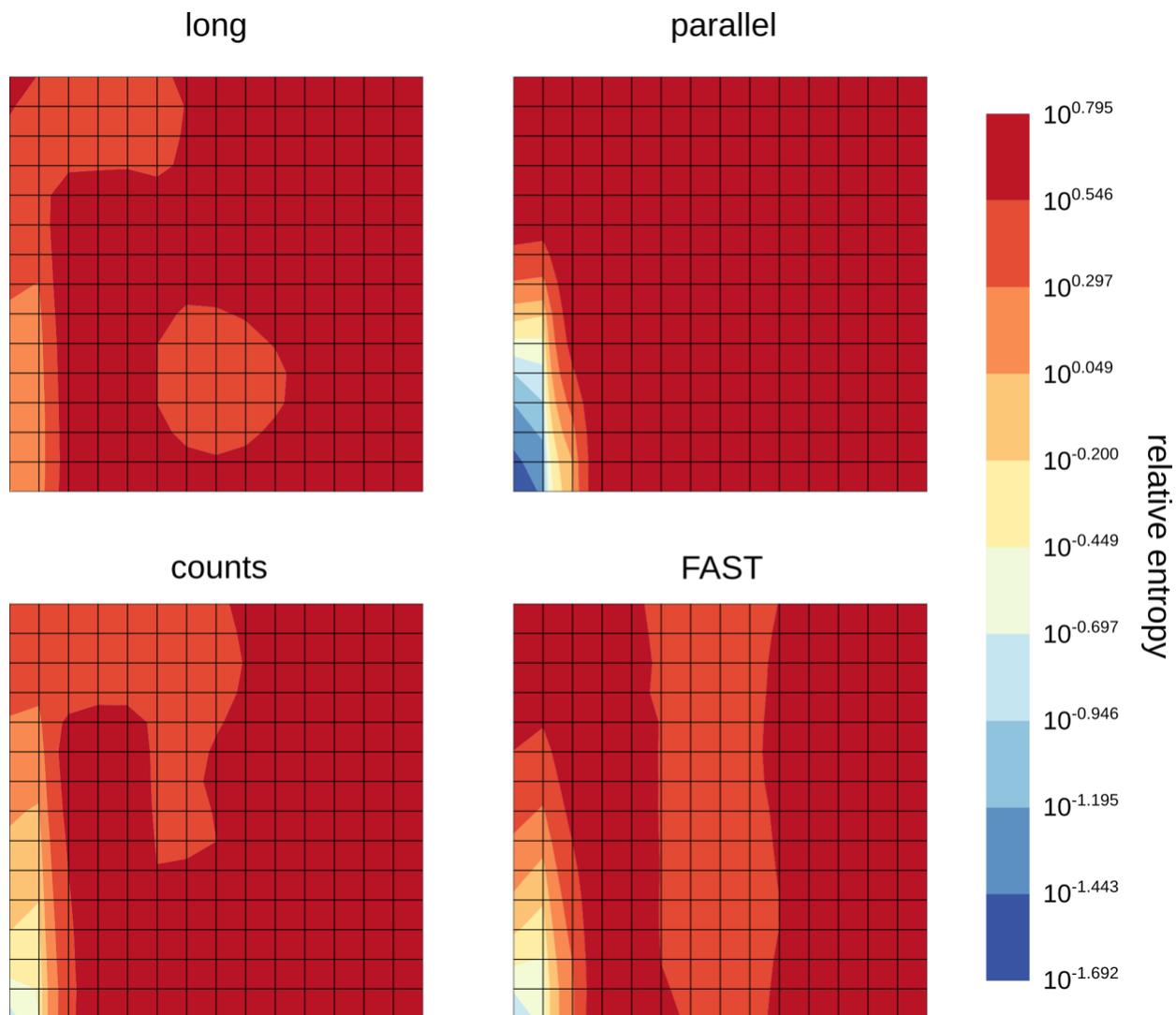

Figure S8: The average Kullbeck-Liebler divergence of each states conditional transition probabilities to the true transition probabilities for the large barriered landscape in Figure 6. Shown are the average divergences of each state for four sampling strategies, a single long simulation, many parallel simulations, counts-based adaptive sampling, and the goal-oriented FAST simulations.

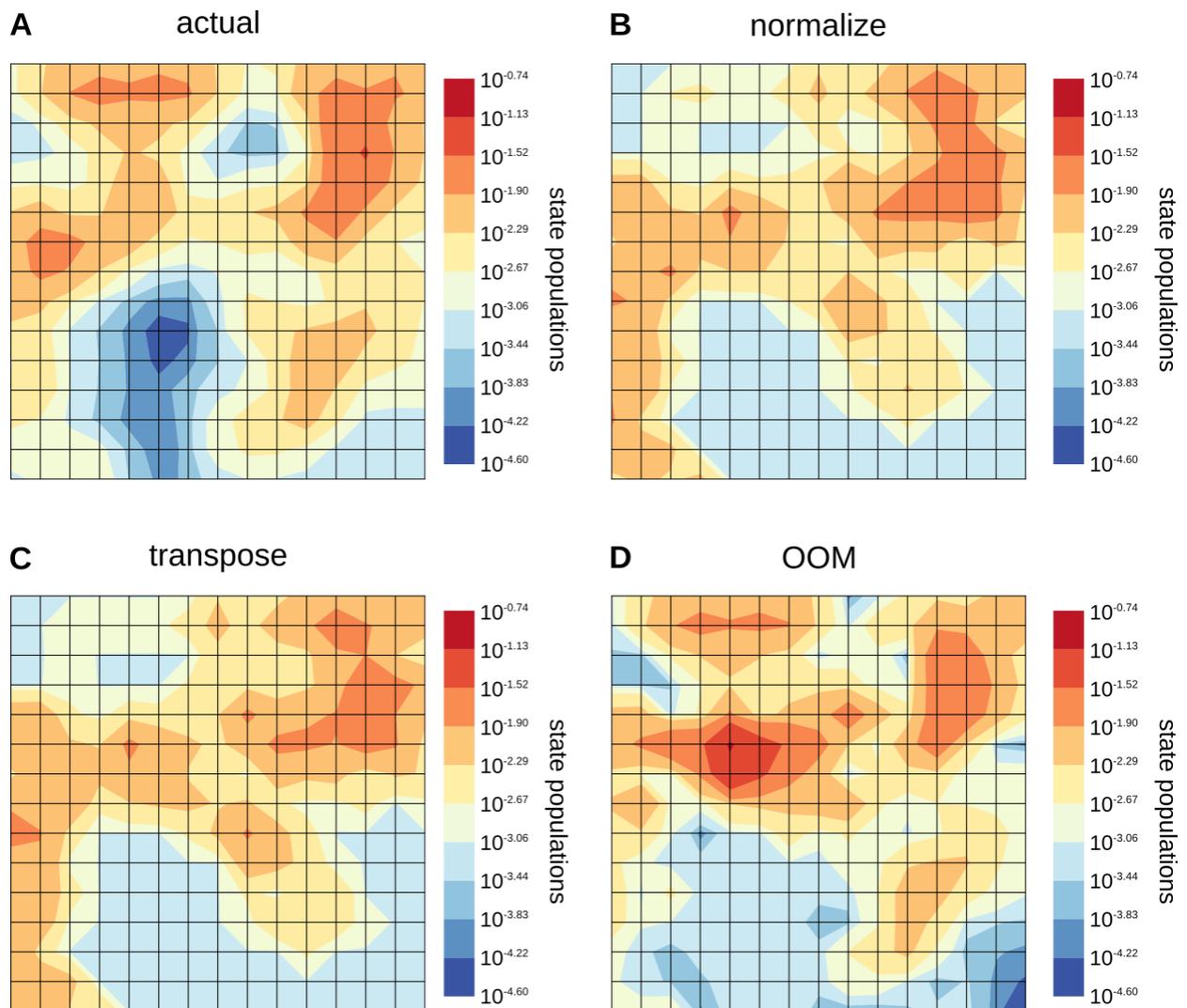

Figure S9: Comparison of MSM estimators' prediction of state populations for a single FAST simulation set. The data set used is the same as is shown in Figure 10A. Shown are (A) the true populations of each state at equilibrium, (B) the predictions from the normalize method, (C) the predictions from the transpose method, and (D) the predictions from the OOM method.

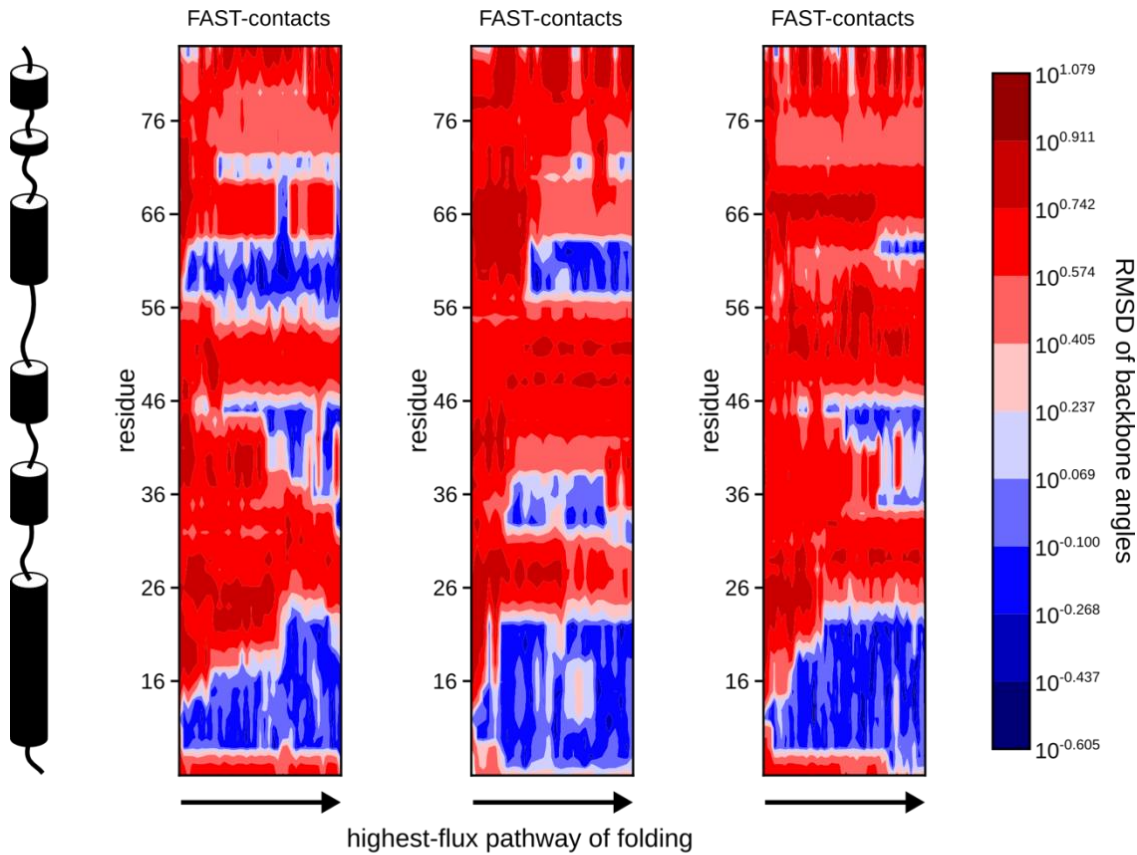
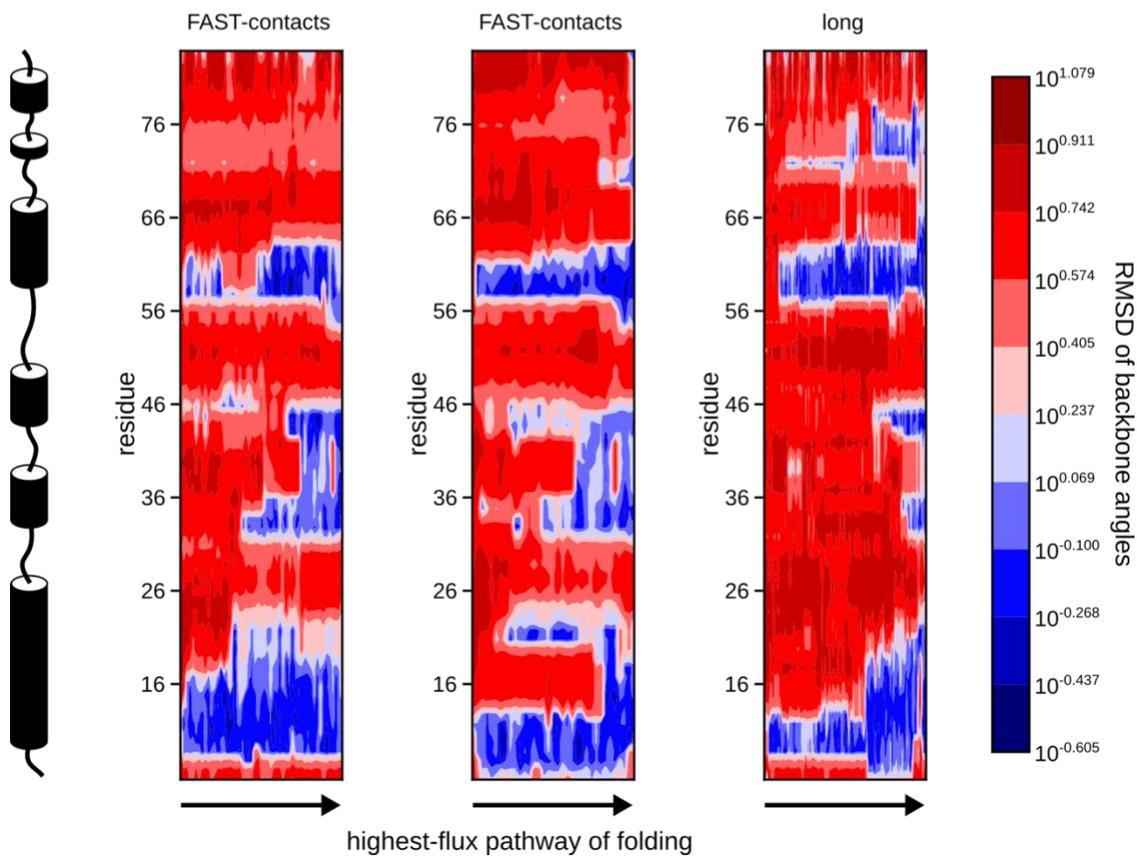

Figure S10: Analysis of λ-repressor predicted folding pathways using the RMSD of each residues' backbone $\phi$ and $\psi$ angles to the crystal structure (PDBID: 1LMB). Folding pathways are determined as an MSMs highest-flux path from the starting state to the state with the largest fraction of native contacts. From left to right on each plot are the residue backbone RMSDs for each state in the predicted folding pathway from five separate runs of FAST-contacts and a single set of long simulations.

Table S1: Probabilities of discovering the target state, average number of states discovered, and relative entropies of transition probabilities for long, parallel, counts, and FAST simulations. Results are shown for 3 landscapes in the main text: 1) funneled landscape depicted in Figure 3, 2) the random barriered landscape depicted in Figure 5A, and 3) the large barrier depicted in Figure 6. Standard deviations of the discover probabilities come from bootstrapping the kinetic Monte Carlo simulations. The discover probabilities for parallel simulations on the funneled and random barriered landscape come from Equation 6 and do not have a calculated standard deviation, since none of these simulations observed a transition to the target state. The optimal value for a given parameter and landscape is bolded.

| Landscape/method | Probability of discovering the target state | Number of states discovered | Relative entropy | Relative entropy of highest-flux paths |
|---|---|---|---|---|
| Funneled | | | | |
| Long | 0.94 ± 3.2E-3 | 144.2 ± 24.0 | 2.53 ± 1.82 | 0.84 ± 0.80 |
| Parallel | 2.2E-5 | 72.7 ± 10.1 | 5.38 ± 0.19 | 2.46 ± 0.05 |
| Counts | 0.62 ± 6.9E-3 | **183.3 ± 12.3** | 4.18 ± 1.74 | 1.96 ± 0.76 |
| FAST | **1.0 ± 7.4E-4** | 168.5 ± 12.3 | **2.02 ± 1.19** | **0.58 ± 0.46** |
| Random barriers | | | | |
| Long | 0.50 ± 7.0E-3 | 108.9 ± 24.7 | 3.15 ± 2.17 | 1.46 ± 1.29 |
| Parallel | 8.5E-7 | 50.2 ± 8.6 | 5.03 ± 0.36 | 2.64 ± 0.22 |
| Counts | 0.34 ± 6.6E-3 | 141.7 ± 18.6 | 3.60 ± 1.69 | 1.89 ± 0.92 |
| FAST | **0.91 ± 4.0E-3** | **143.5 ± 15.7** | **2.61 ± 1.62** | **0.89 ± 0.89** |
| Large barrier | | | | |
| Long | 0.74 ± 5.9E-3 | 129.7 ± 34.1 | 3.69 ± 2.04 | 0.67 ± 0.36 |
| Parallel | 0.059 ± 3.3E-3 | 33.0 ± 7.0 | 5.30 ± 0.20 | 0.83 ± 0.03 |
| Counts | 0.78 ± 5.6E-3 | 146.1 ± 18.7 | 3.97 ± 1.64 | 0.63 ± 0.26 |
| FAST | **0.90 ± 4.3E-3** | 124.8 ± 24.2 | 3.60 ± 1.69 | 0.63 ± 0.29 |
| FAST + string | **0.90 ± 4.3E-3** | **160.6 ± 25.1** | **2.67 ± 1.83** | **0.46 ± 0.32** |